\documentclass[12pt, preprint]{aastex}
\usepackage{graphicx}

\received{} \accepted{}

\shortauthors{Y. P. Wang} \shorttitle{Multi-band evolution of dusty starburst galaxies}

\slugcomment{to submit to Astrophysical Journal}

\begin{document}

\title{Modelling the evolution of dusty starburst galaxies in multi-band deep surveys}

\author{Y. P. Wang$^{1,2}$}
\affil{$^1$Purple Mountain Observatory, Chinese Academy of Sciences, Nanjing}
\affil{$^2$National Astronomical Observatory of China, Chinese Academy of Sciences, Beijing}
\email{ypwang@pmo.ac.cn}

\begin{abstract}
We model the constraints set on the evolution of dusty starburst galaxies by the current deep
extragalactic surveys performed in the far-infrared with $\it Spitzer$, and at radio wavelengths with the VLA.
Our models fit the number counts in all the available spectral bands well, and also provide
a reasonably close match to the redshift distribution of the $\it Spitzer$ detections.
We find: 1.) dusty starburst galaxies with infrared burst phases triggered by galactic interactions at redshift
$z \sim 1-2$ are good candidates to fit the $\it Spitzer$ results
at $24\mu m$, $70 \mu m$ and $160 \mu m$, assuming plausible strengths for the PAH features for the infrared luminous
sources. An Arp220-like spectral energy distribution (SED) for Ultraluminous Infrared Galaxies (ULIGs)
of $L_{\rm ir}>10^{12}\,L_{\odot}$ and one like that of M82 for Luminous Infrared Galaxies (LIGs) of
$L_{\rm ir}\sim 10^{11-12}\,L_{\odot}$ give a successful fit to the $\it Spitzer$ $24\mu m$ and
$\rm ISOCAM\,\,15\mu m$
number counts at flux levels of $S_{\nu}< 1\,mJy$; 2.) the strong evolution of the
number density
of the ULIGs from redshift $z\sim 0$ to
$\sim 1$ predicted by our models is consistent with the current deep
$\rm 1.4\,GHz$ radio surveys and accounts for the upturn in the $\rm 1.4\,GHz$ differential
counts at the sub-mJy flux level; and 3.) comparing with number counts at near infrared bands, as well as the background
measurements using DIRBE and 2MASS, shows that only a fraction of the stellar mass in the
Universe is included in our models of dusty starburst mergers at $z \sim 1 - 2$.

\end{abstract}


\section{Introduction}
During the last few years, remarkable progress has been made in
probing the distant Universe. This progress is driven in part by the
availability of powerful new instruments at various wavelengths spanning the X-ray,
near-, mid- and far-infrared, and submillimeter as well as out to the radio, which
have complemented the traditional studies of galaxy evolution
at $z \ge 1$ based on UV/optical observations.

A crucial advance is the discovery by far-infrared and
submillimetre observations of a significant population of dusty starburst galaxies at
high redshift with infrared luminosities
$L_{\rm ir} > 10^{12}L_{\odot}$, which are ULIGs with 
central activities (AGN or starburst). These objects are heavily hidden by dust
extinction and are triggered either by violent mergers between gas
rich objects or by encounters of galaxies \citep{Sma97, Hug98, Bla99, Eal00, 
Hol99, Pug99, Dol00,
Hug00, San00, Saj03, Xu04}.
The formation rates for massive stars ($\rm M > 5\,M_{\odot}$) 
are approximately
$200-900\,M_{\odot}\,yr^{-1}$, assuming the bolometric
luminosities of ULIGs are dominated by
star formation activities \citep{Con92, Sco97, Ivi98, Ivi00}.

Another important clue to galaxy evolution is the
population of optically faint, near-infrared bright Extremely Red Objects (EROs) often
found as counterparts of sources identified in many other wavebands
\citep{Ric99, Sma99, Ale01, Cha01, Cow01, Lut01, Pie01, Smi01, Ale02, Fra02, Ivi02, Mai02, Pag02, Ste03}.
The classifications of this population are divided between dusty starbursts and more evolved
objects. Morphological
studies by Hubble Space Telescope (HST) and
ground-based near-infrared (NIR) imaging show that active dusty
EROs are mostly disturbed systems, indicating mergers or galactic interactions
could be the driving mechanisms for their behavior \citep{Sma02b, Gil03, Mou04}.
The de-reddened star formation rates in these objects are in a
range of $20 \sim 200\,M_{\odot}yr^{-1}$, with far-infrared luminosities generally
below $ 10^{12}\,L_{\odot}$. In these cases, near-infrared
deep surveys, with detection sensitivity at least three magnitudes
deeper than that of far-infrared and submillimeter surveys, can provide a
possibility to unveil the high-z
dusty starburst galaxies in a way complementary to the surveys at
FIR/submillimeter, which are only sensitive to the very bright infrared sources
\citep{Ken98, Cim99, Cim02, Sma02, Bou05}.

NASA's new infrared facility $\it Spitzer$, which brings significant improvements in IR sensitivity
and survey efficiency over the previous deep infrared cosmological surveys, has produced deep new multi-wavelength
number counts from $24\mu m$ to $160 \mu m$ that
provides a unique opportunity for galaxy evolution studies. Because of their sensitivity to the prominent PAH spectral features
(Polycyclic Aromatic Hydrocarbons), $\it Spitzer$ as well as the Infrared Space
Observatory (ISO) $15\mu m$ deep cosmological surveys can provide an unbiased view on the evolution of
both ULIGs and
LIGs at the redshift range $0<z<2.5$ \citep{Ega04, Le04, Mar04}. Although the initial phenomenological models
fitted the unified ISO/$\it Spitzer$ number counts well, they have been shown not to fit
the source redshift distribution (P\'erez-Gonz\'alez et al. 2005). This
paper uses models based on a binary aggregation description of galaxy evolution
to fit the new data and to provide a more complete test of our understanding of infrared galaxy evolution.

In Section 2 of the paper, we will review our previous work. Section 3 discusses
the main features of the binary aggregation dynamics and the numerical simulation approach we use,
which is centered on an infrared burst phase triggered by gas-rich
mergers. A brief discussion of the construction of template spectra of dusty starburst
galaxies is also given in this section. The model fitting to the recent $\it Spitzer$ results, as well as other
available multi-band deep surveys
spanning from near-infrared to radio
wavelengths, and the estimation of the integrated background
intensities are all presented in Section 4. In section 5, we
summarize our conclusions. The cosmological parameters  $\rm H_{0}=70\,km/s/Mpc,
\,\Omega=0.3$ and $\rm \Lambda=0.7$ are adopted throughout.

\section{Summary of Previous Work}

Theoretical modelling of the evolution of dusty starburst galaxies
constrained by both the
observed Cosmic Infrared Background and the number
counts in far-infrared and submillimeter deep
surveys by IRAS, ISO, and SCUBA indicates a dramatic increase in the number 
density of LIGs and ULIGs towards redshift $z \sim 1$
\citep{Dol00, Wan00, Cha01, Fra01, Chap02, Tot02, Wan02, Lag03}. Wang \& Biermann (2000) and Wang (2002)
discussed the effects of galaxy mergers on the strong evolution seen in these deep surveys
in the context of a
binary aggregation evolutionary scheme. The model requires that the infrared behavior 
be luminosity- and redshift-dependent with a
significant population
of dusty starburst galaxies or
AGNs from gas rich mergers at redshift $z\sim 1 - 2$, and less luminous gas-poor
objects at lower z. The infrared emission of more
massive merging systems is taken to be enhanced
to a higher level and to fade away faster than less massive systems within the
merger time scale of a given epoch. This hypothesis
is based on the observation that ULIGs are usually
more than a factor of 20 brighter than normal starburst galaxies and it is consistent with the recent
proposal for a ``downsizing formation scenario''.

The mechanism driving the strong evolution of infrared luminous sources is still unclear.
It could be, for example, an increment of major/ minor merger rates,
an upturn in the probability of minor tidal
interactions, or an enrichment of cold gas toward high redshift.
Although visual morphological classifications of merger pairs is still very challenging, work
by Bell et al. (2005) and Shi et al. (2005) shows that a decrease of the major-merger rate since redshift $z \sim 1$
is probably not the dominant cause for the rapid decline in the cosmic star formation rate. This finding supports
the prediction
by Wang \& Biermann (2000),
that sub-mJy sharp upturn in
the far-infrared number counts diagram could not be interpreted solely by a merger rate decrease with
cosmic time, but required a luminosity-dependent infrared burst phase from gas rich 
mergers at high redshift and gas poor
mergers at low redshift.
We hope that future sophisticated morphological studies from deep imaging as well as CO mass determination
from ALMA will allow such issues to be addressed more completely.

Although the details of the infrared burst phase are still unclear, it is believed to be related to a
stage of the merger process when the dust mass and
temperature are both dramatically increased \citep{Kle87, Tan98}. Numerical simulations on
the evolution of dusty starburst galaxies by Bekki \& Shioya
(2001) show that there is very strong photometric evolution
during the merger process of two gas-rich disks, and a dramatic
change of the SED around a cosmic
time scale $T \sim 1.3\,$ Gyr, when the two merging disks
become very close and suffer from violent relaxation. The star
formation activities at this moment may reach a maximal rate of $\sim 400\,M_{\odot}$yr$^{-1}$. The
infrared flux in this case increases by an order of magnitude,
especially in the far-infrared ($60\,\mu m
\,\sim 170\,\mu m$) in the emitting frame.

However, such dramatic changes only occur for major mergers. Episodic
starburst phases are indicated by recent observations \citep{Shi05},
and they may tend to smooth out the extreme maxima predicted in the models.
Our model is not
restricted to major mergers. However, for simplicity, we assume in
the current work that galactic interactions would trigger only
a single infrared burst within a merger time scale and simulate a range of such interactions
with a luminosity-dependent infrared burst phase in addition to the
mass-light scaling relation of normal starburst galaxies. We believe that episodic bursts
would not change the main results of this work, but will relax the requirements for a high fraction of
mergers at high redshift. Thus, the current work should
be indicative of the overall pattern of infrared galaxy evolution
regardless of the detailed time dependence of the star formation.

The redshift distribution of ULIGs in the model of Wang
\& Biermann (2000) and Wang (2002) shows a remarkable increase of the number density
from the local Universe up to redshift $z\sim 1$ and a mild decrease towards still higher
redshifts. We have found a similar result based on the study of the cosmic star
formation history using X-ray deep surveys (Wang et al. 2003). This paper examines whether
the evolution of the ULIGs observed in the IRAS and ISO deep far-infrared
surveys is still compatible with recent observations
such as: 1.) in the $\rm 1.4 ~GHz$ radio band, which is itself a star formation indicator, and is an
important way to study the evolution of starburst galaxies; and
2.) $24\mu m$, $70\mu m$ and $160\mu m$ observations by
$\it Spitzer $, which are particularly useful for studying the evolution of the starburst population in the
redshift range $1<z<2$. Another important rationale for this work is to try to learn more about the
relevant physical processes. Although near-infrared number counts, unlike mid- or far-infrared counts,
would not provide strong constraints on the star formation activities in our modelling,
we include here a comparison of our modelling with the results of near-infrared deep surveys,
to get an estimate of the fraction of stellar mass contained in these
starburst merging galaxies to the total stellar mass budget in the Universe.

With model constraints ranging from near-infrared out to deep radio surveys, we have constructed an overall model for both ULIGs and LIGs that evolves in a continuous way,
based on a merger triggered starburst scenario. This approach differs from other models that assume ULIGs
evolve separately from LIGs and with an extremely strong evolutionary rate.

\section{Model Description and numerical simulation}

\subsection{Key Numerical Relations}

We describe the modelling algorithm briefly in this section. Readers are referred to Wang \& Biermann (2000) and
references therein for
the details of the binary aggregation dynamics and the numerical techniques. An aggregation phenomenon
based on the
Smoluchowski (1916) equation is adopted in the model to trace the
evolution of the mass distribution function with cosmic time, which is in the form of:

\begin{eqnarray}
\label{von}
\frac{\partial\rm N(m,t)}{\partial t}=\frac{1}{2}\int_{0}^{m}dm^{'}K(m^{'},m-m^{'},t)
N(m^{'},t)N(m-m^{'},t)-N(m,t)\int_{0}^{\infty}dm^{'}K(m,m^{'},t)N(m^{'},t)
\end{eqnarray}

\noindent $N(m,t)$ is the mass distribution function in the
``comoving'' frame.
The kernel $K(m,m^{'},t)=n_{g}(t)<\Sigma V(t)>$ reflects the
interaction rate for each pair of masses $(m, m^{'})$, and
depends on the mechanism and environment of such
aggregations. $n_{g}$ is the density of galaxies; $V(t)$ is the relative velocity of the
interacting pairs, and $\Sigma$ is the cross section. In aggregation dynamics, the interaction kernel $K =
n_{g}<\Sigma V>$ is the key point for driving the whole
evolutionary process, which depends strongly on the
environmental structures. However, the determination of the interaction kernel suffers from various
uncertainties and
complexities. We shall adopt in this model a simplified
formula for the aggregation kernel $K(m,m^{'},t)$ with
separated time evolution and cross section
terms as below:

\begin{equation}
\label{kmk}
K(m,m^{'},t) \propto
t^{k}(m^{2/3}+{m^{'}}^{2/3})\left[1+\alpha(m^{2/3}+{m^{'}}^{2/3})\right].
\end{equation}

\noindent where $t^{k}$ reflects the evolution of the aggregation rate with cosmic time. The free
parameter $k$ depends on the specific structures and interacting environments. A range of values is discussed by
Cavaliere \& Menci (1997) for groups, clusters and large scale structures. The second free parameter is $\alpha$,
which describes the relative importance of
two kinds of encounters (geometric collisions and focused resonant interactions) in the cross section.
The aggregation time scale $\tau$ is inversely proportional to
the interaction kernel $K(m,m^{'},t)$ i.e., $K(m,m^{'},t) \propto \tau^{-1}$.

A Monte Carlo inverse-cascading process is used to simulate
the binary aggregation of galaxies
described by Eq. (\ref{von}). This equation has been solved in a number of cases \citep{Saf63, Tru71,
Sil78}, which show that the dynamical process does not strongly depend on
the initial conditions, i.e. the mass spectrum is independent of initial details with
self-similar evolution. For this study, we adopted a
$\delta$-function at $M=2.5 \cdot 10^{9} M_{\odot}$ as an initial mass spectrum (so dwarfs are included in the
new calculation) and start the simulation from redshift $z=15$. In the previous studies, we had
set the initial mass to $M=2.5 \cdot 10^{10} M_{\odot}$.
Other model parameters of the dynamical processes are discussed in detail by Wang \& Biermann (2000)
and are still valid for the current modelling.

Up to now, the mass distribution
function $N(m,t)$ could be derived from the Smoluchowski equation using the Monte Carlo simulation. However,
to compare the modelling with various observational constraints,
such as luminosity functions of
galaxies in general or with the morphologies, source counts, redshift distributions and background
intensities from multi-band surveys, we need to understand the conversion of the mass distribution function
$N(m,t)$ to the
observable luminosity function $N(L,t)$, as well as the modelling of a
luminosity-dependent infrared
burst phase from gas-rich mergers, all of which are described in detail by Wang
\& Biermann (1998, 2000) and Wang (1999, 2002). We summarize several main points as follows.

A simple prescription of the mass-light ratio for the faint blue
galaxies is given by Cavaliere \& Menci (1997) as $L/L_{\ast} =
(M/M_{\ast})^\eta$ ($L_{\ast}$
is the local standard characteristic luminosity with the corresponding mass
$M_{\ast}$). Here, $\eta = 4/3$ if the cross section is
purely geometrical; this value is consistent with the observational 
results by Kormendy (1990) . $L_{\ast}\propto f(z,\lambda_{0},\Omega_{0})$ can be used to describe redshift dimming
or luminosity evolution. Simplifying the color and K-
corrections, we get roughly that $L_{B}\propto
\frac{L_{\ast}}{M_{\ast}^{\eta}}\,M^{\eta}$. We assume in
our model an infrared to optical color ratio $\frac{L_{60\mu m}}{L_{B}}\propto
M^{\acute{\eta}}$ ($\acute{\eta}\sim 1.5$ is adopted in the calculation). This value is consistent with both the
numerical calculations of the photometric and spectroscopic evolution of dusty starburst galaxies, and with the
current understanding
of the ULIGs, which are believed to be the
extremely luminous infrared burst phase due to starburst merger
events where the far-infrared luminosity, $L_{\rm ir}$, is enhanced both by
accumulation of dust and by an increase in the dust
temperature. This burst phase can enhance the infrared
luminosity by a factor of about 20 over that of normal starburst
galaxies \citep{Kle87, Sil98, Tan98, Bek01}. We adopt $L_{60\mu m}\propto
\frac{L_{\ast}(0)}{M_{\ast}^{\eta}}\,f(z)\,M^{\eta+\acute{\eta}}$,
$f(z)\propto (1+z)^{\beta}$ in the modelling. A value of $\beta \sim 5$ after a transition redshift
$z\sim 1$ gives the best fit, which successfully interprets the sharp upturn at the faint
end of far-infrared
number counts. In addition, choosing $z\sim 1$ as a transition redshift, with gas rich mergers at $z \ge 1$
and gas poor mergers at $z < 1$, is also consistent with the cosmic time scale
at $z\sim 1$, which is about $3\times 10^9$ years, approximately the time scale for disk evolution.
Galaxies may become gas poor during this stage. The scaling factor of the mass-light ratio is
normalized by the local luminosity function from the IRAS survey.

Except for understanding the physical basis for the two major free parameters in
the interaction kernel ($\rm k \sim -1.35$ and $\alpha \sim 1$),
which we adjust
in the simulation to give a good fit to the current observations, the uncertainties in the relative emission
over the radiation spectrum across cosmic time as galaxies evolve, including progressive evolution and episodic
starburst activities triggered by galactic interactions etc., are now the major challenges to galaxy
evolution studies. With only number counts, redshift distributions, and integrated background levels, which
are rather loose constraints, we simply assume a luminosity dependent infrared burst phase
after a transition
redshift, to mimic the change of the SED shapes (especially in the far-infrared)
for galaxies with different luminosities.
We simulate such an effect by adding a scaling factor $(\frac{L_{\rm ir}}{L_{\rm c}})^{-\zeta}$ to the luminosity
evolution term defined in the last paragraph ($f(z)\propto (1+z)^{\beta}$), where $L_{\rm c}$ is the minimal
luminosity bin of
the calculated luminosity function. The best fit result gives $\zeta \sim 0.9$ at far infrared wavelengths.

This luminosity dependent infrared burst reflects a physical reality, that the infrared
luminous galaxies at the bright tail of the luminosity function
become gas poor faster than the less luminous ones and fade away quickly within the merger time scale of that
epoch. This behavior is consistent with the current concept of a ``downsizing formation scenario''.
In addition, it indicates that less luminous infrared sources, such as LIGs, could have infrared burst phases
persisting longer than those of ULIGs. In fact, we can describe such an idea
with an infrared burst ``duty circle,"
$\rm f_{on}$. This term is based on expressing the duration of the infrared burst phase
in terms of the merger time scale,  $\rm t_{sb}/t_{merger}$,
where $\rm t_{sb}$ is the starburst time scale and $\rm t_{merger}$ the merging time scale. Learning from
observations, we have $\rm t_{merger}\propto (1+z)^{-m}$ and $\rm t_{sb}\propto \frac{1}{L_{ir}^n}$. Therefore,
we can derive an empirical relation $\rm f_{on}\propto \frac{1}{L_{ir}^n}\,(1+z)^m$. From the discussion above,
we now understand that this luminosity dependent infrared burst phase is actually implicitly related to the
starburst time scale.

\subsection{Model Characteristics}

While the basic concepts of our model are clearly plausible, 
the details of the necessary assumptions are somewhat arbitrary. However,
it seems that they are very important for fitting the sharp upturn at faint flux levels in the
current deep far-infrared surveys.
For a detailed comparision with the color behavior vs. flux densitiy,
we shall in the future
include a real physical model including both proper chemical evolution and dust emission
properties.

In the current calculations, starburst galaxies, dust shrouded AGNs 
and ``non-evolving'' galaxies are treated separately as three major
classes of infrared emitting sources. The ``non-evolving'' galaxies represent the big spiral and
elliptical galaxies, which were already in place about 8 Gyr ago and are
considered to have a fixed luminosity function with cosmic time \citep{Lil98, Sch99}. We know from recent work by
Hammer et al. (2005) that some intermediate-mass spiral galaxies may have experienced starburst
phases due to galactic interactions, which are counted within the starburst population triggered by galactic
mergers/interactions in our calculation. Although the treatment of the three populations separately in the
modelling is a bit arbitrary, this simplification should not affect the current results significantly.

Unlike other models that assume ULIGs evolve separately from LIGs and with an extremely strong evolutionary rate,
we prefer to think that merger-triggered
infrared luminous sources including both ULIGs and LIGs exhibit a continuous
evolution, where the infrared emission of ULIGs may
change more rapidly than that of less luminous sources (i.e. LIGs may have a longer infrared burst duration than
that of ULIGs). We
simulate such a continuous range of evolution by assuming a
luminosity dependent infrared burst phase based on the consideration of different time scales for the starburst
galaxies of different infrared luminosities to consume the
cool gas. The AGN component is included as an additional model constraint, since the dust enshrouded geometry
could result in significant radiation from them at infrared wavelengths. We assume AGN
activities are triggered by galactic mergers or interactions based on the aggregation dynamics, similar to the
starburst population. The statistics of the starburst and AGN populations are normalized by the observed local
luminosity function of starburst galaxies at $60\,\mu m$
from Saunders (1990) and that of Seyferts from Rush et al. (1993). The abundance of dust-shrouded AGNs is very
uncertain, especially for those
dust enshrouded QSOs at high redshift. For the models, we set the abundances of the dust-shrouded AGNs
relative to the total
AGN population to be $50\%$ and
$80\%$ at local and high redshift, respectively, based on the statistics from
Hubble Space Telescope imaging survey of nearby AGNs and the
cosmic X-ray background \citep{Mal98, Gil99}.

\subsection{Template SEDs}

\begin{figure}
\epsscale{0.80}
\plottwo{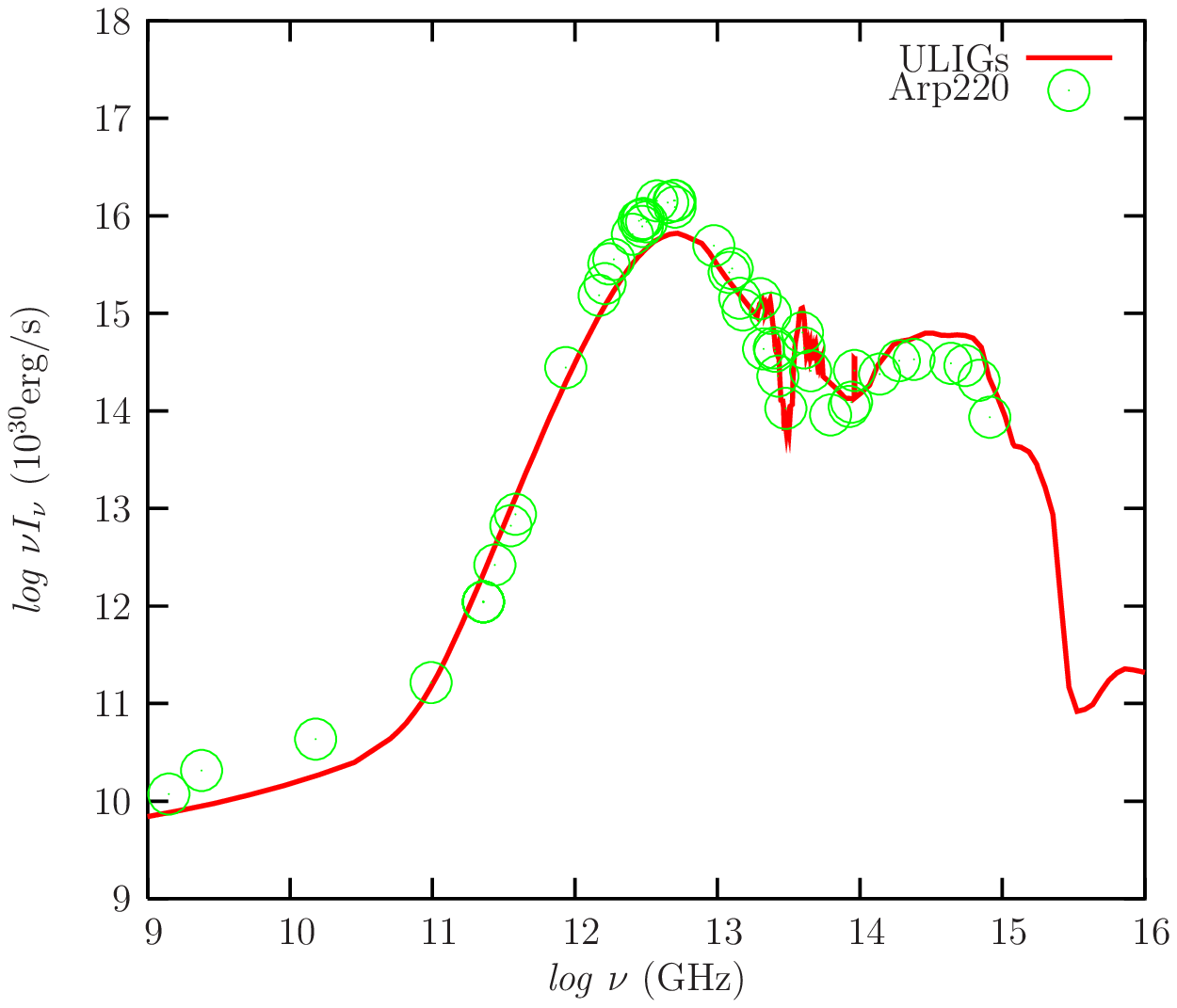}{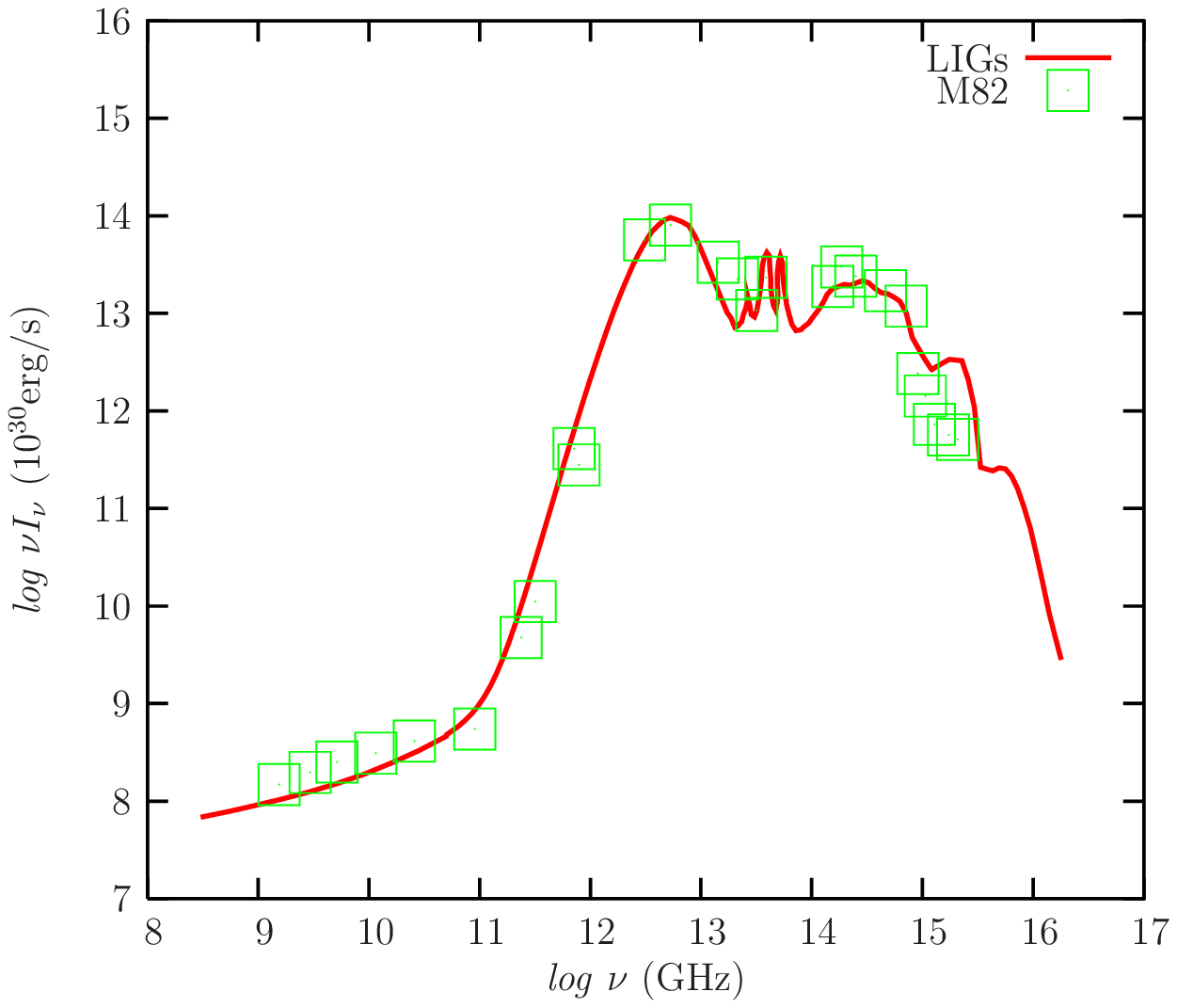}
\caption{Left: The SED of ULIGs selected for the modelling that gives a successful fit for the present deep surveys,
especially in the far-infrared. The data points are photometric measurements of Arp220 (corrected to rest frame),
collected by Elbaz et al. (2002); Right: A comparison of the SED of LIGs selected for the modelling with the
photometric measurements of M82. The data points are collected by Silva et al. (1998).
The meanings of the lines are denoted in the figure.  \label{fig1}}
\end{figure}

To construct a template
spectrum for the high redshift dusty starburst sources around
redshift $z\sim 1$, we utilize the publicly available program
GRASIL introduced by Granato and Silva to produce a library of SEDs.
We replace the PAH features with a template
for PAH emission based on rest-frame $\rm ISOCAM/CVF\,\,5-16.3\mu m$ spectra of Arp220 and
M82 (Weedman et al. 2004). We selected two template SEDs respectively for 
ULIGs and LIGs, which are adequate to fit the present deep surveys, 
especially the $24\mu m$ number counts by $\it Spitzer$ and the ISOCAM $15\mu m$ results.
A comparison of the template
SED for ULIGs and LIGs with the measurements of Arp220 and M82 is given in Fig. \ref{fig1}, which shows that the
dust re-radiation peaks around $50-60\,\mu m$, consistent with a dust
temperature of $\rm T \sim 55\,K$ and a dust absorption coefficient
$K_{\lambda}\propto \lambda^{-\beta}$ with the coefficient index $\beta \simeq 1.2$.
A third template, the SED of M51, is assumed for the spiral population.

An important change has been made in our current modelling, compared with our earlier work (Wang 2002). 
Previously, only a simple template SED (without specific
consideration of PAH features) was
assigned for the whole infrared luminous starburst population, rather than the separate
SEDs for LIGs, and ULIGs. To give a good fit to the new $\it Spitzer$ results, it requires
adjustment of the merger
rate especially at higher redshifts and of the free parameters in the prescription of the
infrared burst phase. This probably contributes significantly to the better fits we achieve for the
redshift distributions of the infrared sources compared with previous efforts
that have assumed a single SED (e.g., Lagache et al. 2004).

We choose the template spectrum for the obscured AGNs at low redshift 
to be that of NGC 1068, and the unobscured ones that of the mean SED
of a Seyfert I. The early phases of these AGNs are assumed to show
typical spectra such as the dust shrouded F10214+4724, and a phase
poor in cold gas like the Cloverleaf quasar. The templates of all these spectra
were modelled by Rowan-Robinson (1992), Rowan-Robinson et al. (1993) and Granato
et al. (1994, 1996, 1997). 

\section{Results and discussion}

\subsection{Mid- and Far-Infrared Number Count Fitting}

\begin{figure}
\plottwo{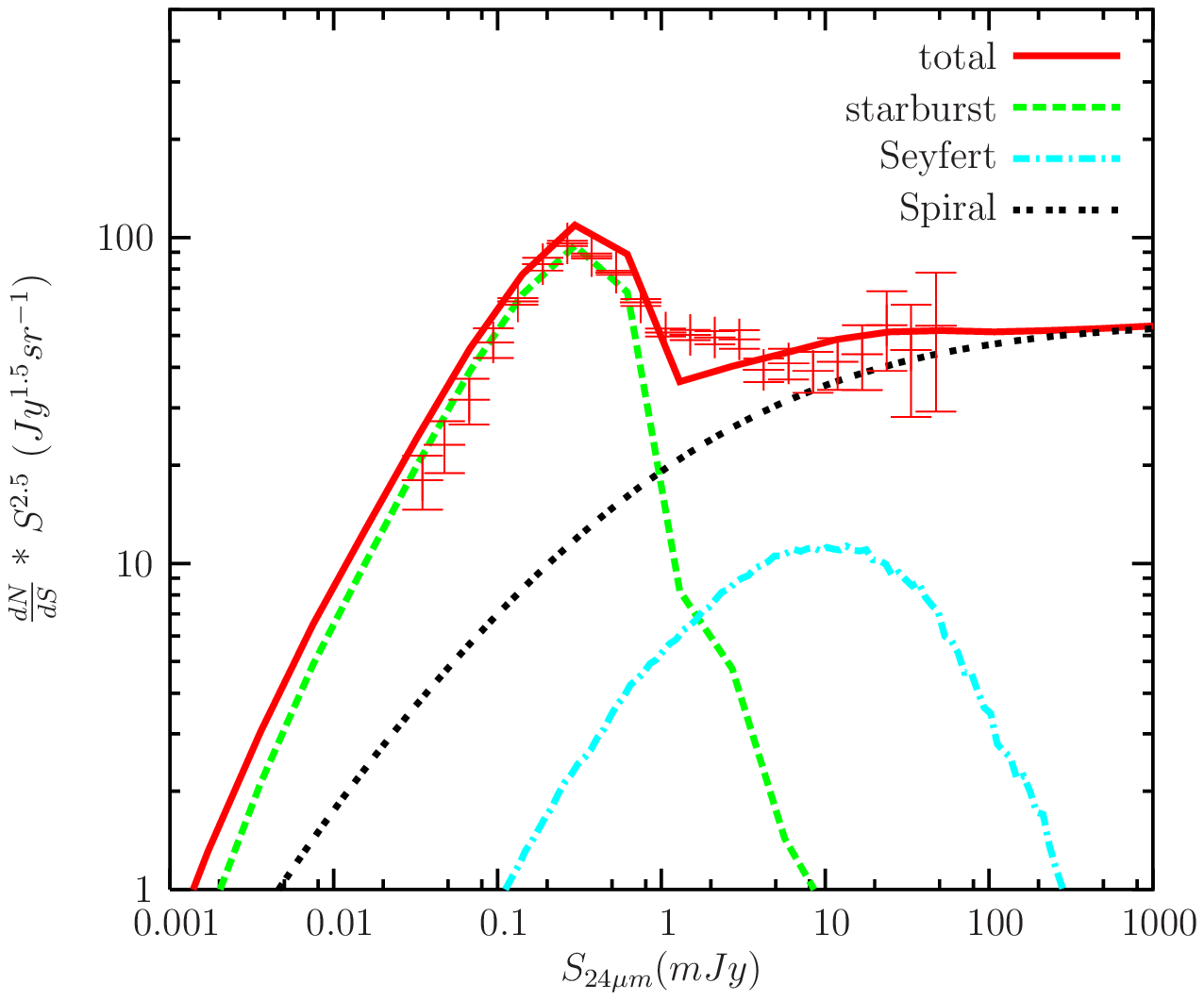}{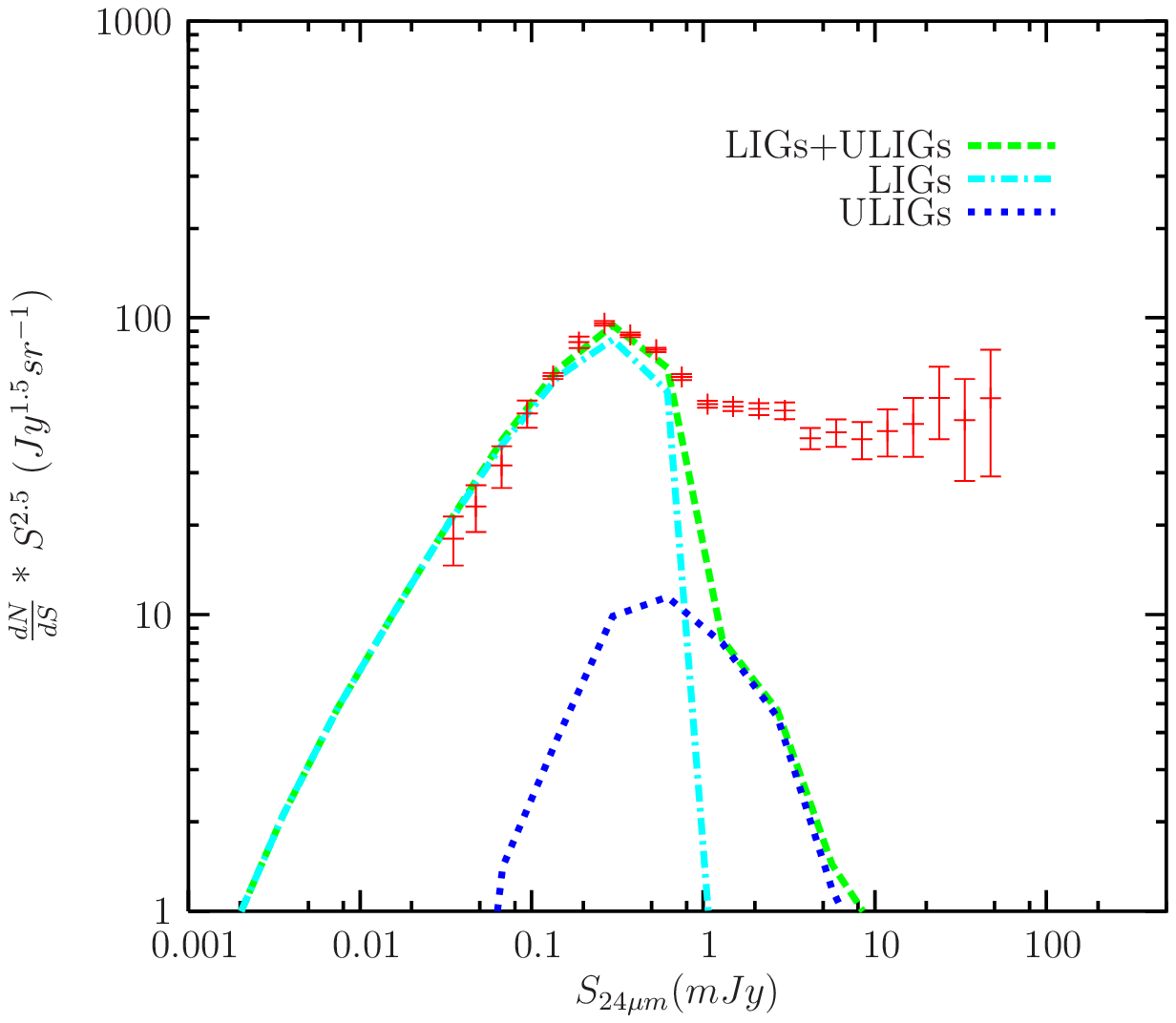}
\caption{Left: The model fitting of the differential number counts in $\it Spitzer$ $24\mu m$ deep surveys.
The data points are
from Papovich et al. 2004, normalized to a Euclidean slope, which show the average counts from all
the $\it Spitzer$ fields and corrected for completeness;
Right: The calculated contribution from LIGs, ULIGs and LIGs+ULIGs to the differential number counts at $24\mu m$.
We see that LIGs dominate the $\it Spitzer$ $24\mu m$ number counts at the sub-mJy flux level, while ULIGs
contribute mostly to the bright part of the peak at $0.3\,mJy <S_{\nu}\,<\,1\,mJy$. \label{fig2}}
\end{figure}

\begin{figure}
\plottwo{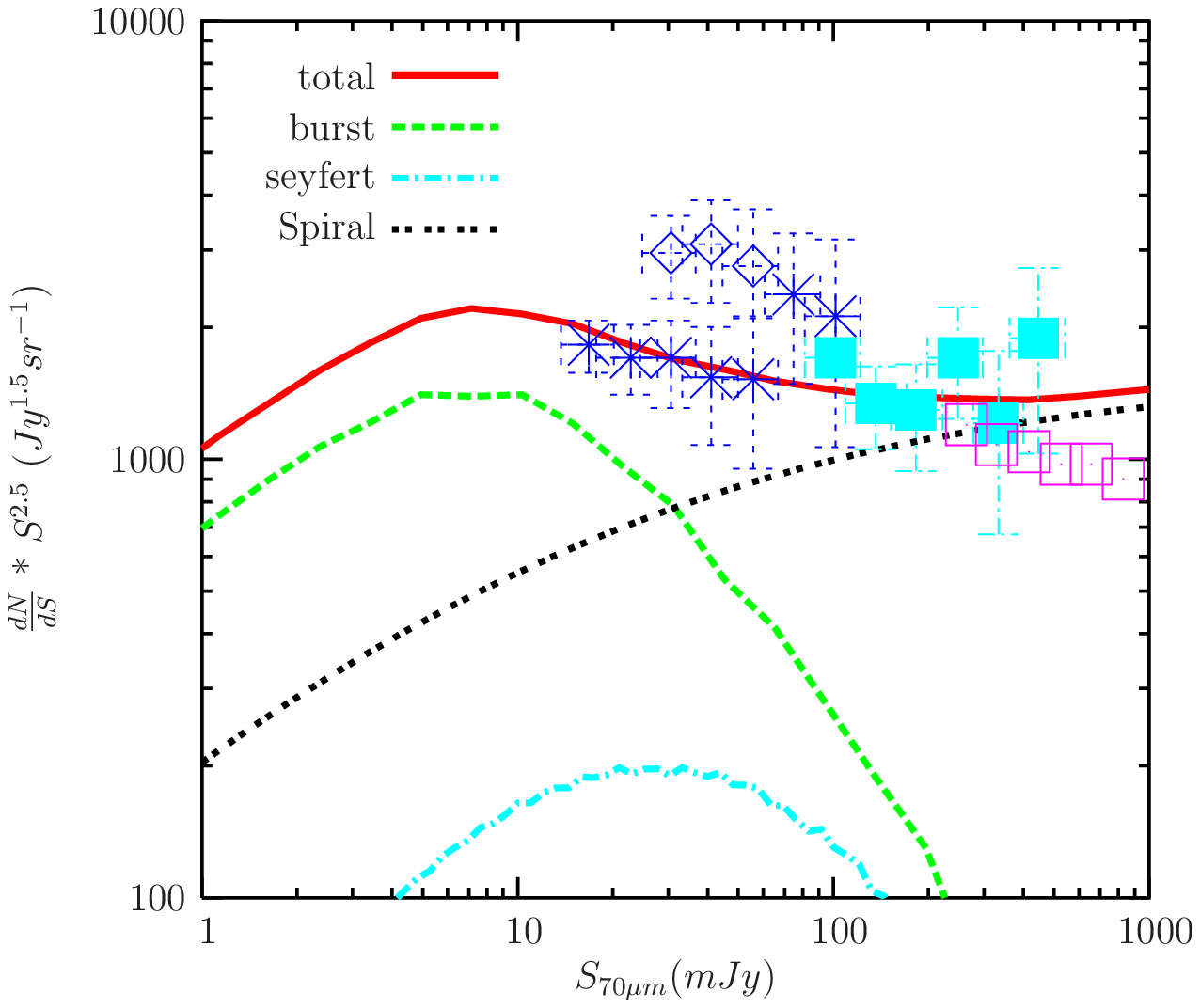}{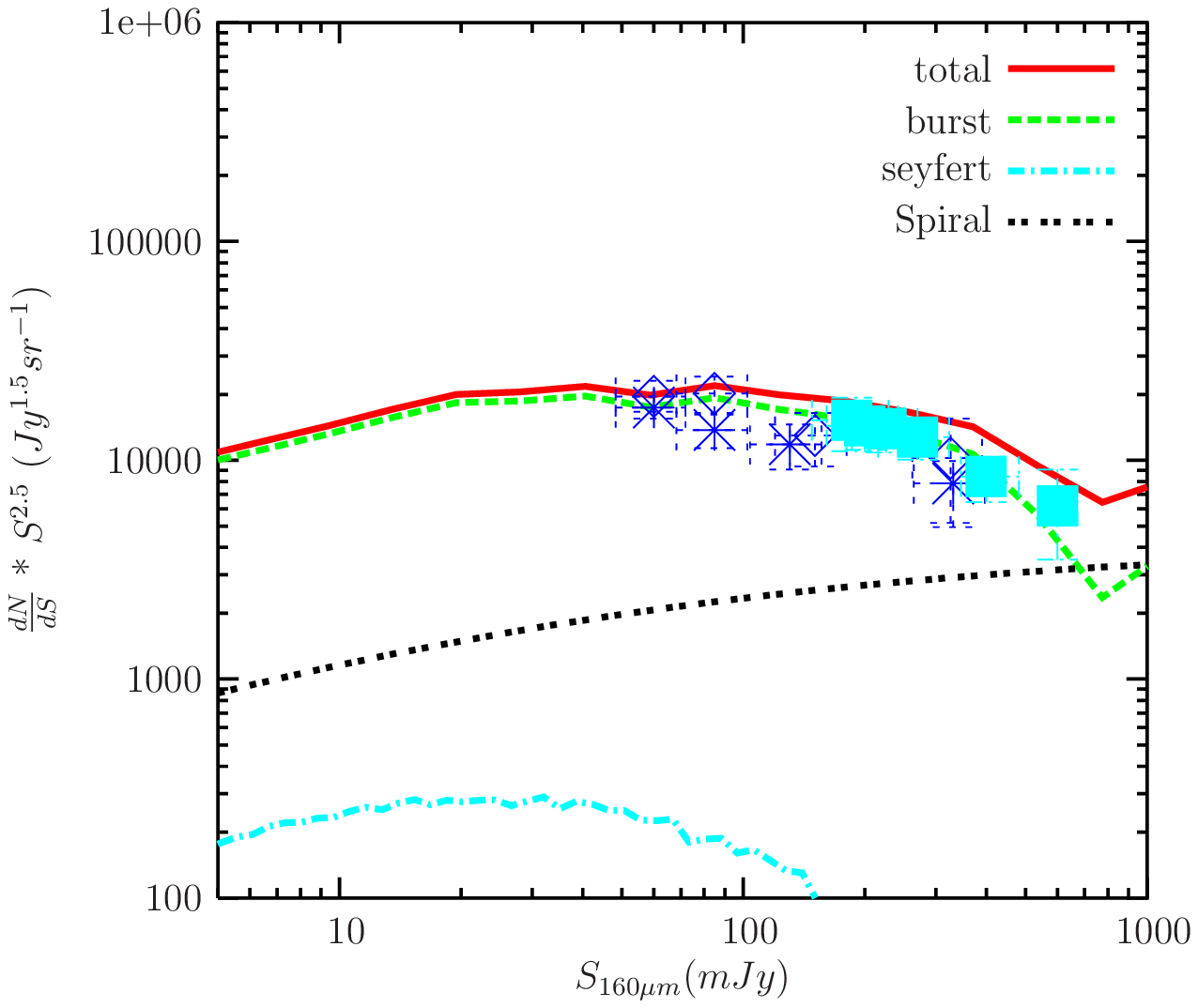}
\caption{Left: The model fitting of the differential number counts at $70\mu m$. The blue asterisks (CDFS),
blue empty diamonds (Marano field), and blue filled squares (Bootes field) are source counts in $\it Spitzer$ deep
surveys with no correction for incompleteness (Dole et al. 2004). Red empty squares are IRAS $60\mu m$ counts from
Lonsdale et al. (1990) converted to $70\mu m$; Right:
The differential number counts fitting at $160\mu m$ by the model calculation. The blue
stars (CDFS), blue empty diamonds (Marano field) are the source counts at $160\mu m$ with no correction for
incompleteness (Dole et al. 2004). Blue filled squares correspond to ISO FIRBACK 170um
number counts (Dole et al. 2001).  \label{fig3}}
\end{figure}

We first show in Fig. \ref{fig2} and Fig. \ref{fig3} the comparison of our model results with
the number counts from recent $\it Spitzer$ MIPS deep surveys at $24\mu m$, $70\mu m$ and
$160 \mu m$. The green dashed line is used to denote the population of
starburst
galaxies with an infrared burst phase due to galactic interactions. The black dot-dotted line
corresponds to the non-evolving galaxies, and the blue dash-dotted line is for Seyferts/AGNs.
The sum of the three
populations is represented by the red line. A similar set of lines will be used in all the following plots.

The left panel of Fig. \ref{fig2} is the model fitting of the differential source counts in $\it Spitzer$
$24\mu m$ deep surveys, which include $\sim 5\times 10^4$ sources to an $80\%$ completeness of $\sim 80\mu Jy$.
$\it Spitzer$'s increased sensitivity and efficiency in large area coverage over previous infrared
facilities, coupled with the encompassment of PAH emission in $\it Spitzer$'s $24\mu m$ band, dramatically
improve the quality and statistics of number counts in the mid-infrared \cite{Pap04}. The right panel of
Fig. \ref{fig2} shows the calculated contribution to the differential number counts at $24\mu m$ 
from ULIGs and LIGs separately, as well as the sum of the two
populations. As shown in Fig. 3, 
we find that the faint end upturn of the number counts at $70\mu m$ and
$160\mu m$ can be interpreted successfully by the
population of starburst mergers from our modelling with infrared luminosities $L_{\rm ir}> 10^{12}\,L_{\odot}$ and
Arp220-like SEDs. 

\begin{figure}
\plottwo{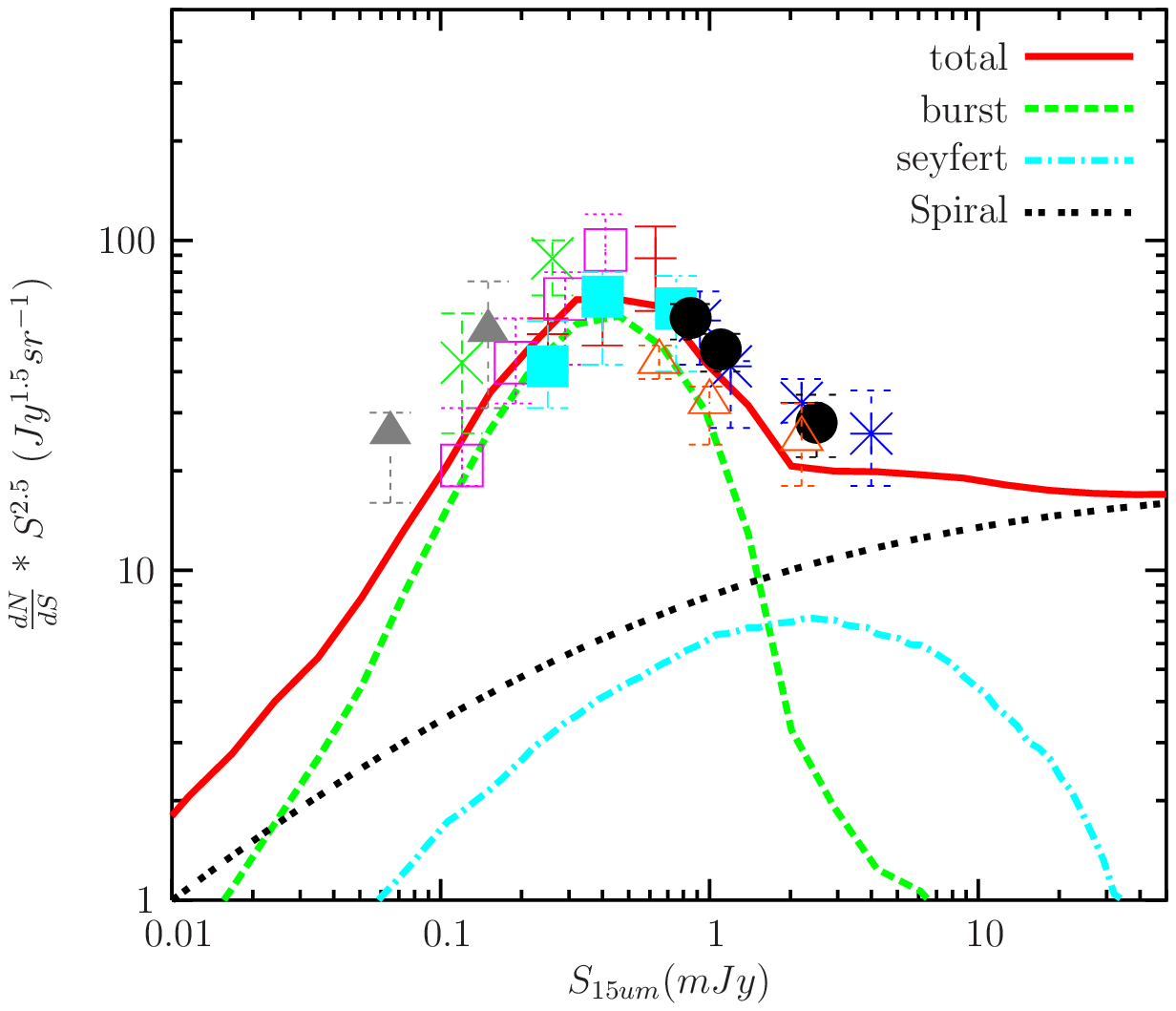}{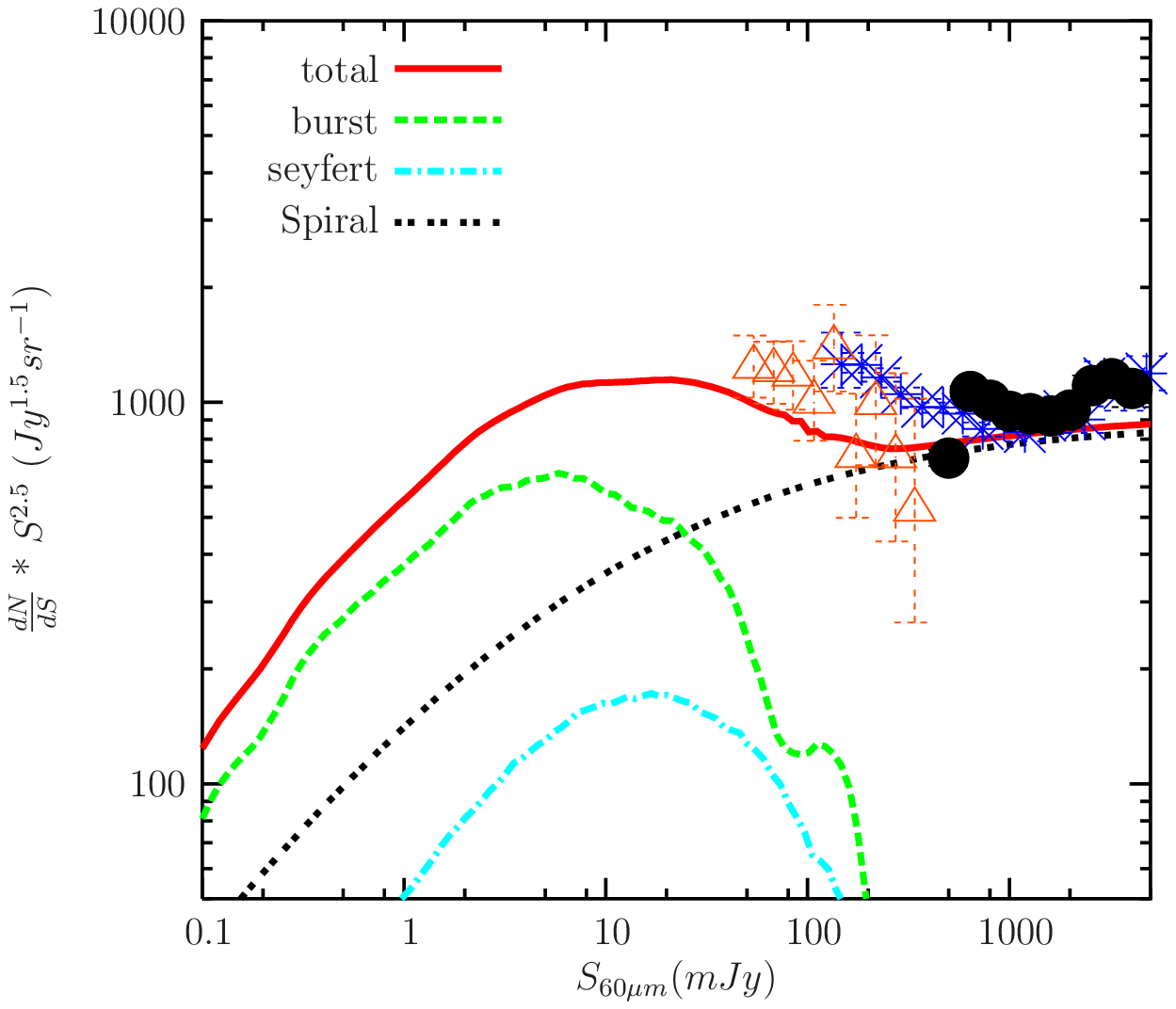} \caption{Left: Model fitting of the
differential number counts at ISOCAM $15\,\mu m$. The data points
are from a variety of ISO deep surveys (Elbaz et al. 1999); Right:
IRAS $60\,\mu m$ number counts fitting. The data points are from
the IRAS Point Source Catalogue (1985)(PSC, black filled circles),
Hacking et al. IRAS deep survey (HCH, red empty triangles), FSC
(blue asterisks) from deep surveys by Moshir et al. (1992) and
Saunders (1990). The kink seen at the bright end of the burst
population number counts is due to the numerical inaccuracy of the
logarithm binning during the Monte-Carlo simulation. \label{fig4}} 
\end{figure}

\begin{figure}
\plottwo{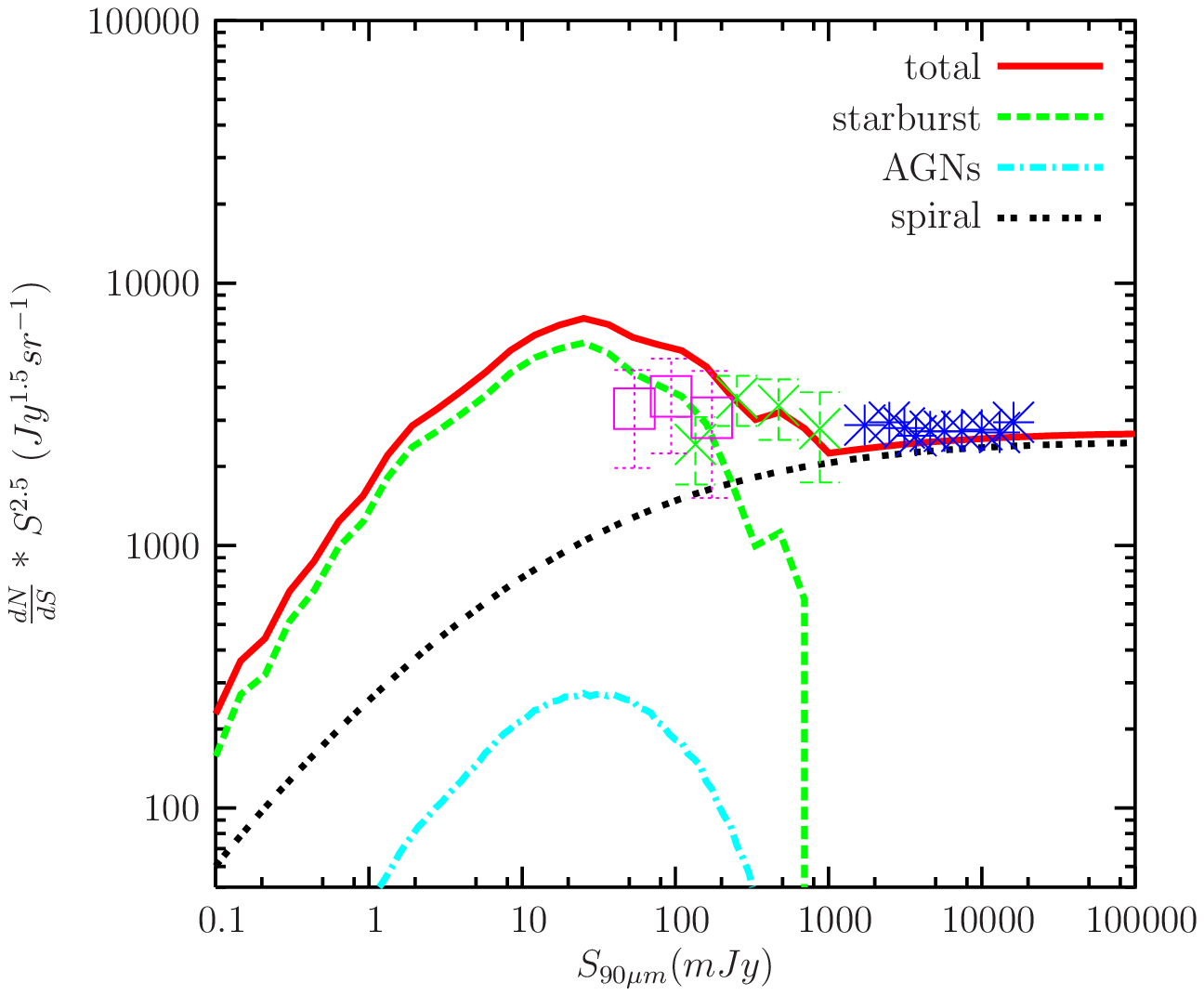}{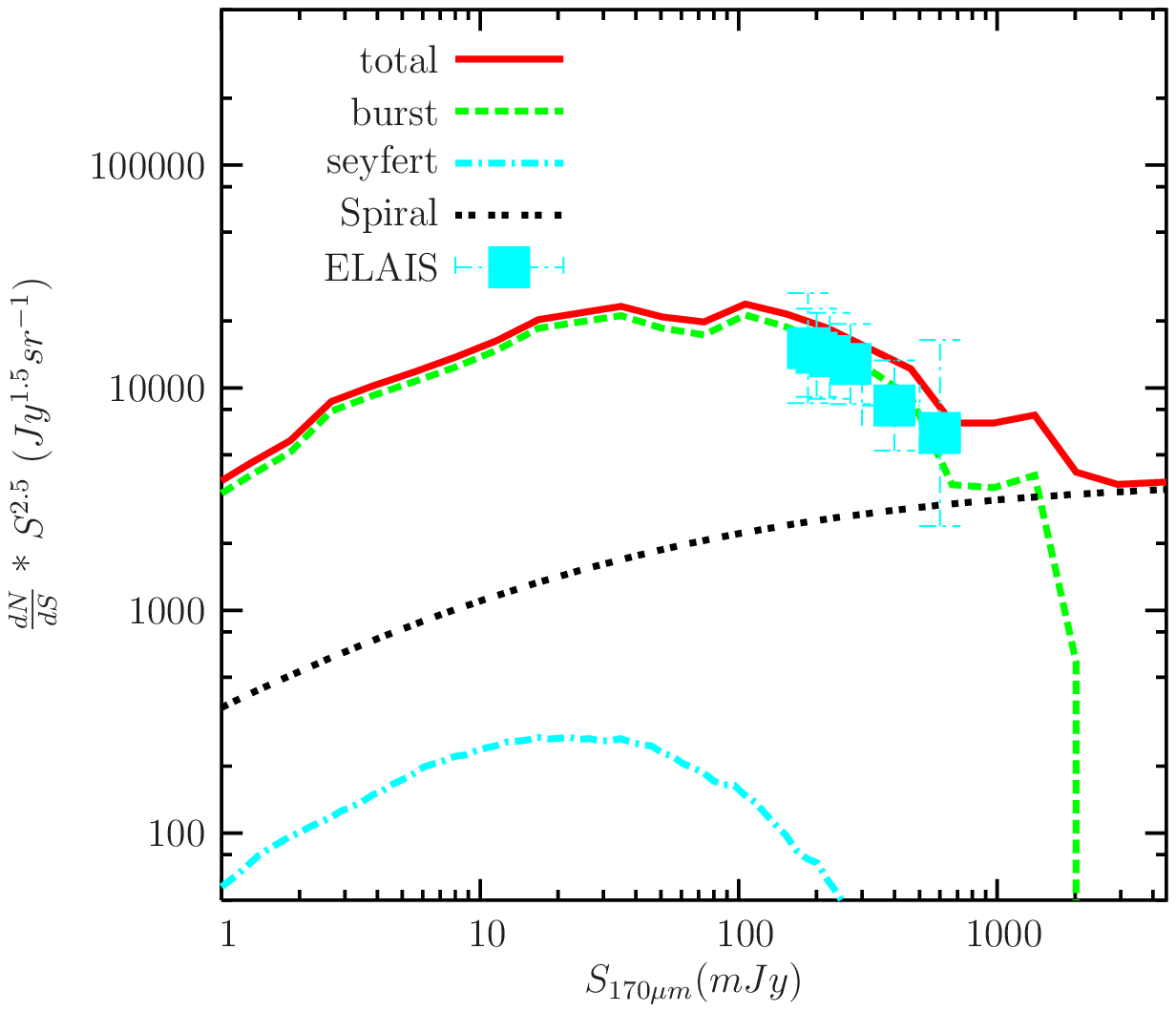} \caption{Left: The model fitting of the
ELAIS normalized differential counts at $90\,\mu m$. The data are
from recent $90\mu m$ final analysis by H$\acute{e}$raudeau
(2004)(empty squares) and Lockman Hole counts from Rodighiero
(2003)(crosses), IRAS counts (stars) for galaxies in PSCz
catalogue; Right: The differential number count fitting for
FIRBACK $170\,\mu m$ deep survey by the model. The data points are
from Dole et al. (2001). The kinks in the modelled number counts
arise for the same reason as in Fig. 4. \label{fig5}}
\end{figure}

\begin{figure}
\plottwo{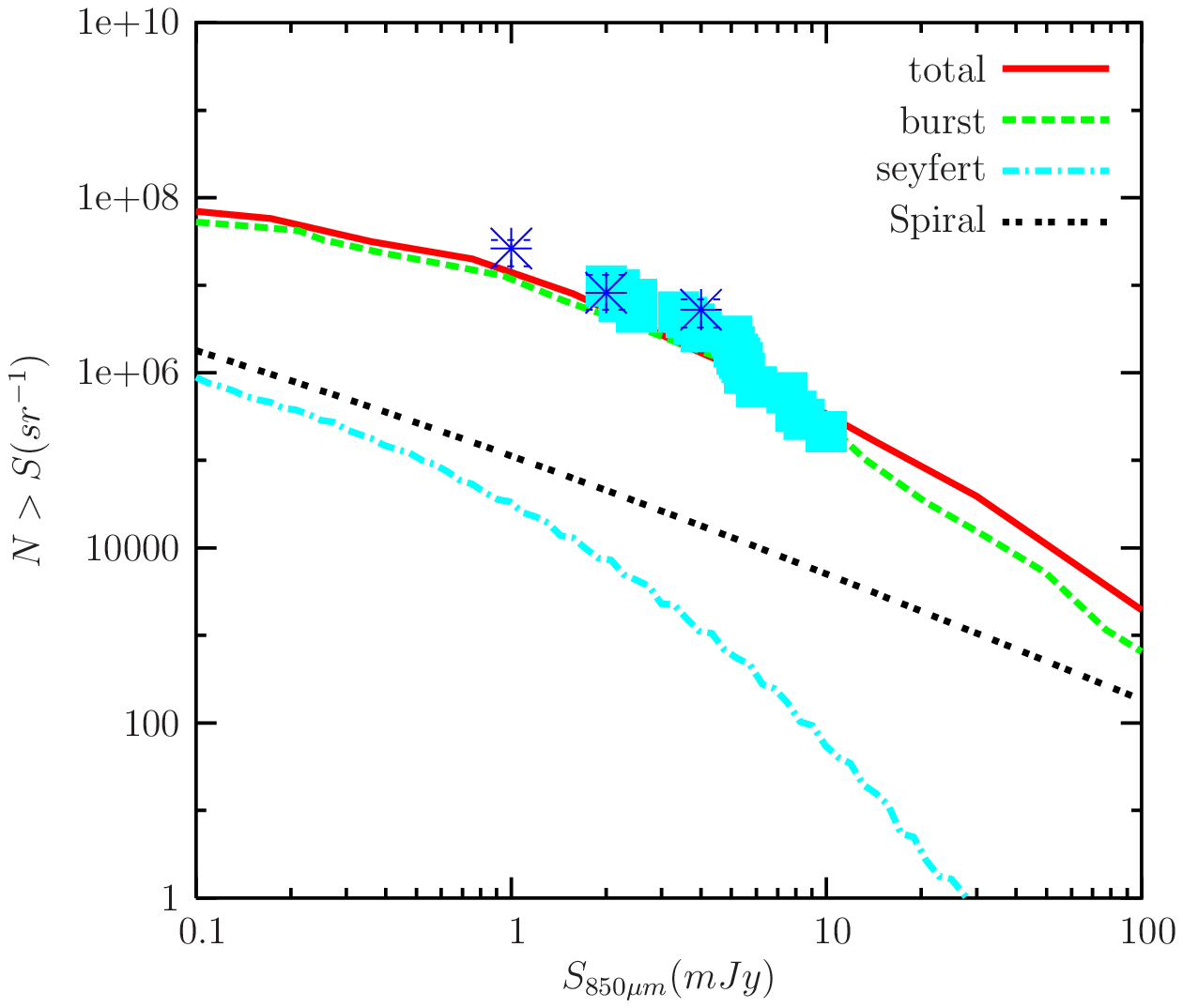}{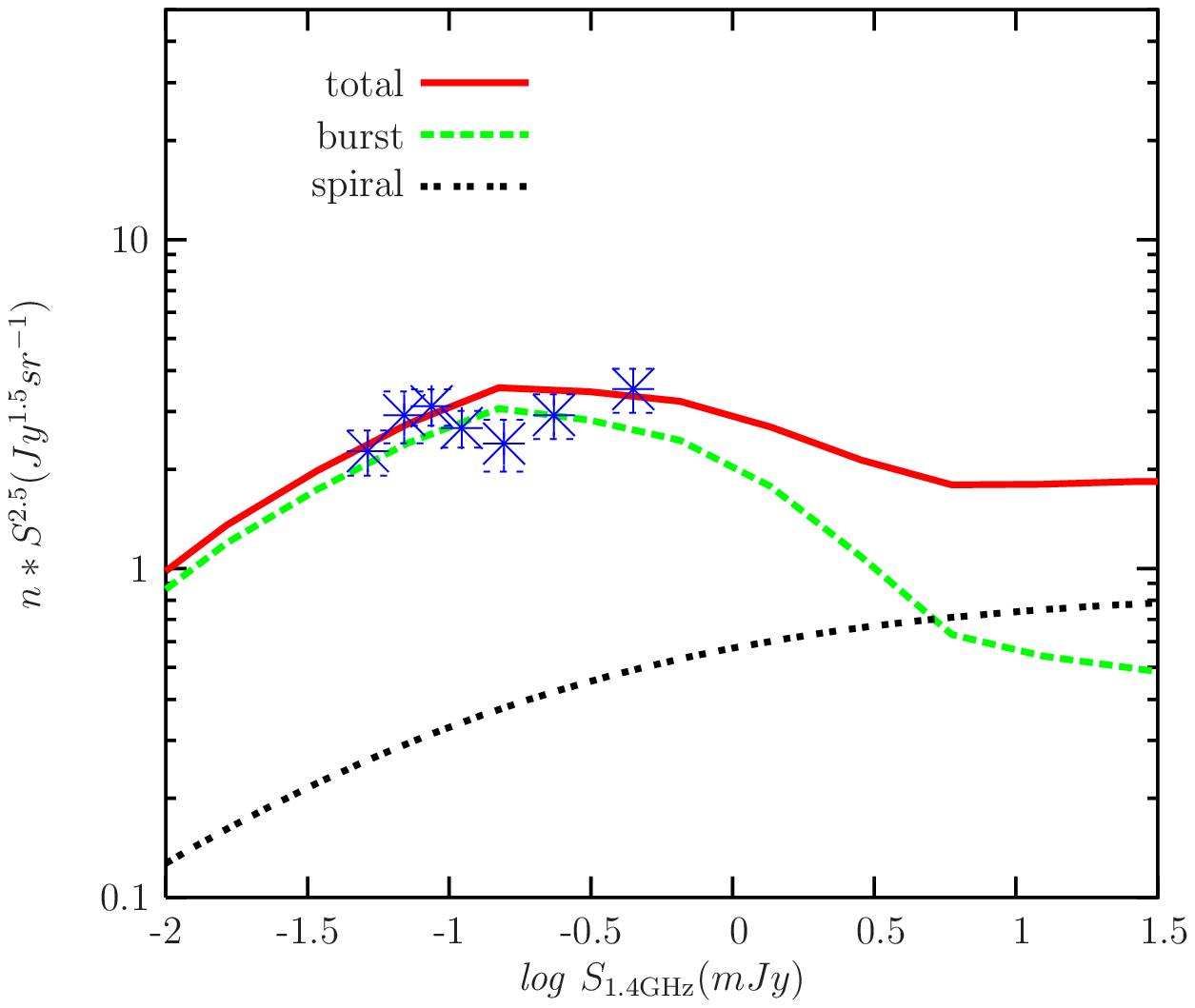} \caption{Left: Model fitting of the
integral number counts at $850\,\mu m$. The blue asterisks are
from Blain et al. (1999), and the blue filled squares are from
Barger et al. (1999); Right: The differential number count fitting
of the $\rm 1.4\,GHz$ radio deep survey by the modelling. The blue
asterisks are the $\rm 1.4\,GHz$ counts in the HDF-N from Richards
(2000). The model result demonstrates that IR emitting starburst
galaxies do not contribute significantly to the $\rm 1.4\,GHz$
counts at the bright end of the number counts diagram, but are
important for the population of radio sources at flux densities $\le
0.5-0.8\,mJy$. The $\rm 1.4\,GHz$ number counts at the bright end
($\ge 1\,mJy$) are primarily due to radio AGNs, which are not
relevant to our discussion here. \label{fig6}}
\end{figure}

Fig. \ref{fig4} and Fig. \ref{fig5} present the model fitting of the number counts from ISOCAM $15\mu m$ deep
survey, IRAS $60\mu m$, ELAIS $90\mu m$ final analysis and FIRBACK $170\mu m$ surveys.
Fig. \ref{fig4} (Right) to Fig. \ref{fig6} (Left)
show that the number counts for a reliable
subset of the detected sources at FIR/Submm wavelengths
could be sufficiently accounted for by the infrared burst phase
when a population of ULIGs with
$L_{ir}\ge 10^{12}\,L_{\odot}$ might be produced by
merger-triggered starburst/AGN activities at $z\sim 1$ \citep{Kaw98, Elb99, Efs00, Dol01}. However, we also
learned from the simulation that the FIR slope of the adopted SEDs would significantly affect the FIR/Submm number
count fitting at faint flux levels, while the free parameter $\zeta $ influences mostly the bright part of the source
count diagram. Similar changes as for
the {\it Spitzer} data were made in the other far infrared number count fittings,
basically for the same reasons. For example, the peaks of the
calculated differential number counts at IRAS $60\mu m$ and ISO $90\mu m$ are shifted to a fainter flux
(from $\sim 50\, mJy$ to $\sim 10\, mJy$) (the exact observational values are still
open for debate due to the detection limits of
current far-infrared surveys). 

The simulation shows that the extremely strong evolution at
sub-mJy of the $15\mu m$ and $24\mu m$ number counts is mainly dominated by LIGs with enhanced PAH emission
(see Fig. \ref{fig2} (Right)), while ULIGs contribute mostly to the bright part of the peak at
$0.3\,mJy < S_{\nu}< 1\,mJy$. The bright end of the $24\mu m$ number counts diagram
is mostly contributed from ``non-evolving'' spiral population and dusty AGNs.
We also see that the PAH spectrum and strong silicate absorption of the ULIGs
affect significantly the contribution of the dusty starburst galaxies to the $\it Spitzer$ $24\mu m$ and
ISO $15\mu m$ number count
diagrams in the flux range $1\,mJy<S_{\nu}<10\,mJy$. This is one of the reasons why the starburst population is
almost negligible above a few mJy in the newly calculated $15\mu m$ number counts, while it was dominant up to
$\sim 10\,mJy$ in the previous work \citep{Wan02}. Of course, there are a few other modifications to
better fit the recent $\it Spitzer$ $24\mu m$ number counts: 1.) we include the dwarf population with initial mass
spectrum $M=2.5\times 10^9\, M_{\odot}$; 2.) we increase the scaling factor of the interaction kernel
$K(m, m^{'}, t)$ in equation (2) to shorten the merger time scale of dwarfs or minor mergers at high redshift to a
reasonable range and enlarge the absolute value of the free parameter $k$ to
make the exponential damping faster; and 3.) the free parameters in the prescription of infrared burst duration
($\beta, \zeta$) are adjusted to change the relative infrared burst duration of LIGs and ULIGs.

\subsection{Number Counts at Other Wavelengths}

The model fitting of the differential radio counts at $\rm 1.4\, GHz$ is shown in Fig. \ref{fig6} (Right). We found
that the well known $\rm 1.4\,GHz$ source excess observed at sub-mJy levels can be
interpreted satisfactorily by the population of starburst galaxies with an infrared burst phase 
at $z \sim 1$. These objects are LIGs of $L_{\rm ir}\sim 10^{11-12}\,L_{\odot}$
with enhanced PAH emission. We believe that the $\rm 1.4\, GHz$ source excess 
and the strong evolution detected by ISOCAM $15\mu m$ and $\it Spitzer$ $24\mu m$ deep surveys at
flux $S_{\nu}< 1\,mJy$ are due to the appearance of the same population.

\begin{figure}
\plottwo{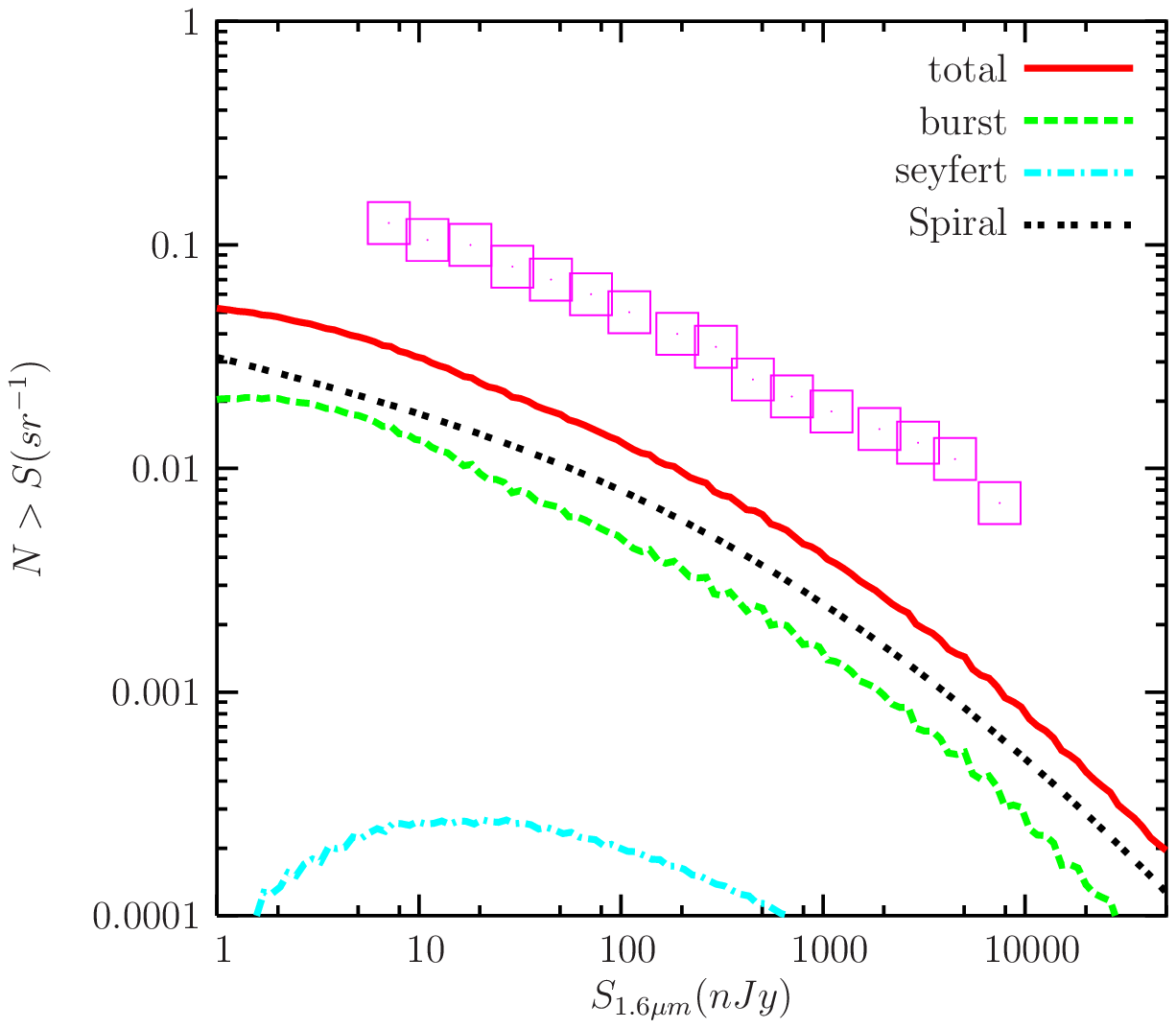}{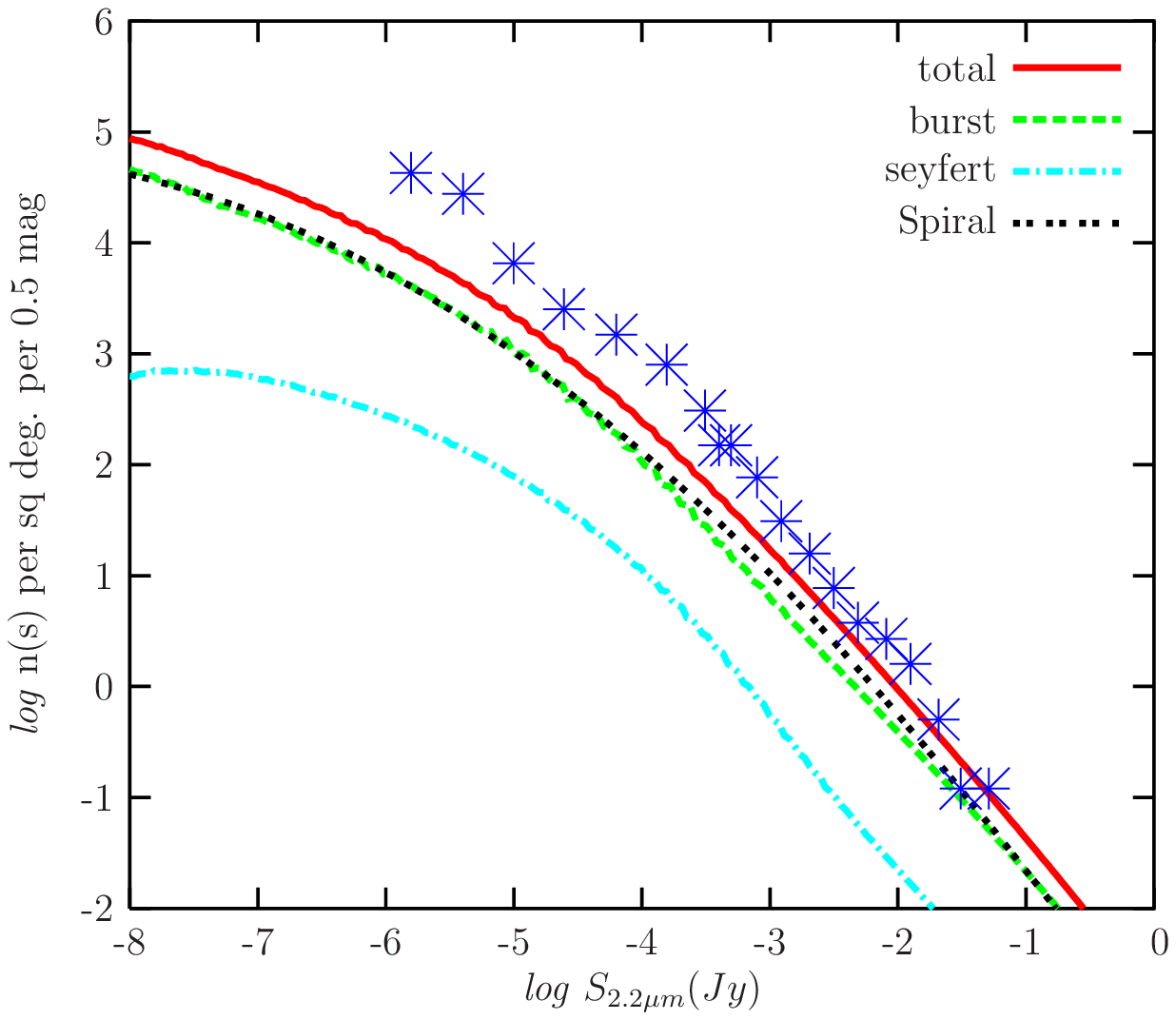}
\caption{Left: Model fitting of the cumulative number counts of infrared sources at $1.6\mu m$ brighter
than a given threshold $F_{\nu}$. The data points are from NICMOS observations
covering $1/8-{th}$ of the Hubble Deep Field North, which 
reaches a flux level about three magnitude fainter than the deep surveys at K-band and ISOCAM $6.7\mu m$
(Thompson et al. 1999); Right: source count fitting by our modelling at $2.2\mu m$. Data are from
McCracken et al. (2000).\label{fig7}}
\end{figure}

\begin{figure}
\epsscale{0.80} \plotone{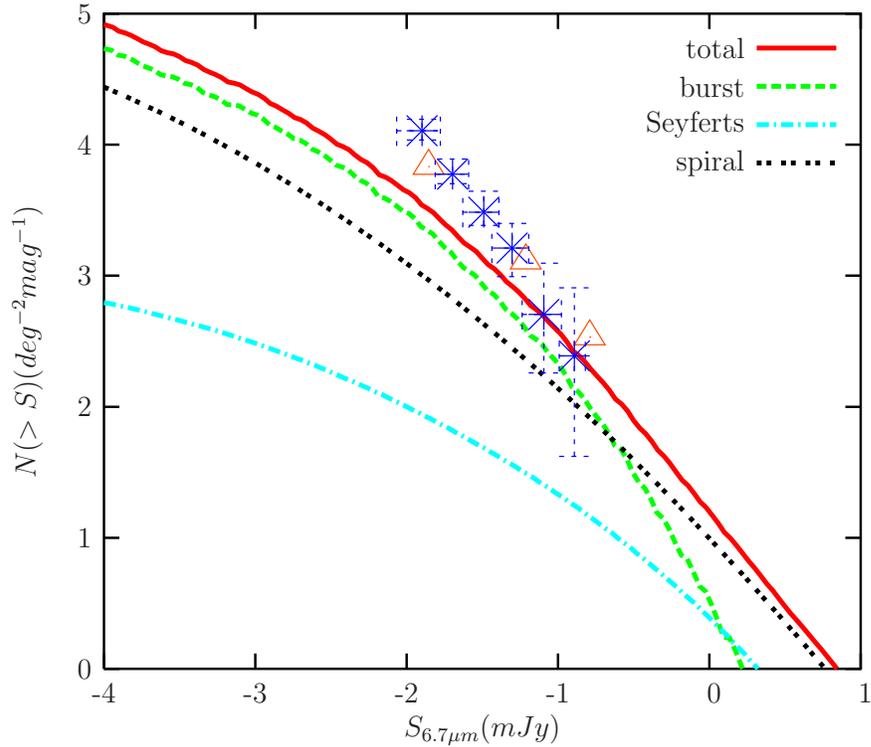} \caption{Model fitting of the
cumulative number counts of infrared sources at $6.7\mu m$ with
$S_{\nu}>10\,\mu Jy$ within $16\,arcmin^{2}$. Data are from
Metcalfe et al. (2003, red empty triangles) and Sato et al.(2003,
blue asterisks). \label{fig8}}
\end{figure}

Fig. \ref{fig7} (Left) shows the model fitting of the cumulative number counts of
galaxies at $1.6 \mu m$ brighter than a given flux threshold $\rm
F_{\mu}$. The data are NICMOS H-band observations covering $1/8$ of
the area of the Hubble Deep Field North that reach a flux level down to $\sim nJy$ (about three magnitudes deeper
than other near- or mid-infrared deep surveys). The number count fittings of the K-band and ISOCAM $6.7\mu m$ deep
surveys are presented in Fig. \ref{fig7} (Right) and Fig. \ref{fig8} respectively.
We know from these results that the population of dusty starburst merging galaxies in the model only contributes
to the bright end of the number counts at K-band and $6.7\mu m$. There is a significant deficiency of
sources from our modelling at the faint
flux levels ($\le 0.01\,mJy$), which is clearly shown in Fig. \ref{fig7} (Left). This indicates that deep near-infrared
observations reveal a population of very faint and high redshift galaxies that are not detected
in current mid-infrared surveys. We estimate roughly that $30\%$ to $50\%$
of the stellar mass in our Universe could be contained in the population of dusty starburst merging galaxies in
our modelling, based on the near-IR number counts fitting and the background intensity.
However, this value should be taken as an upper limit since the near-infrared light and stellar mass
correlation will be biased in these starburst
merging systems by the intense star formation activities, which can result in the near-infrared
being dominated by young supergiants and the galaxies having a much
smaller mass-to-light ratio than ellipticals.

\subsection{Redshift Distribution and Cosmic Background}

\begin{figure}
\plottwo{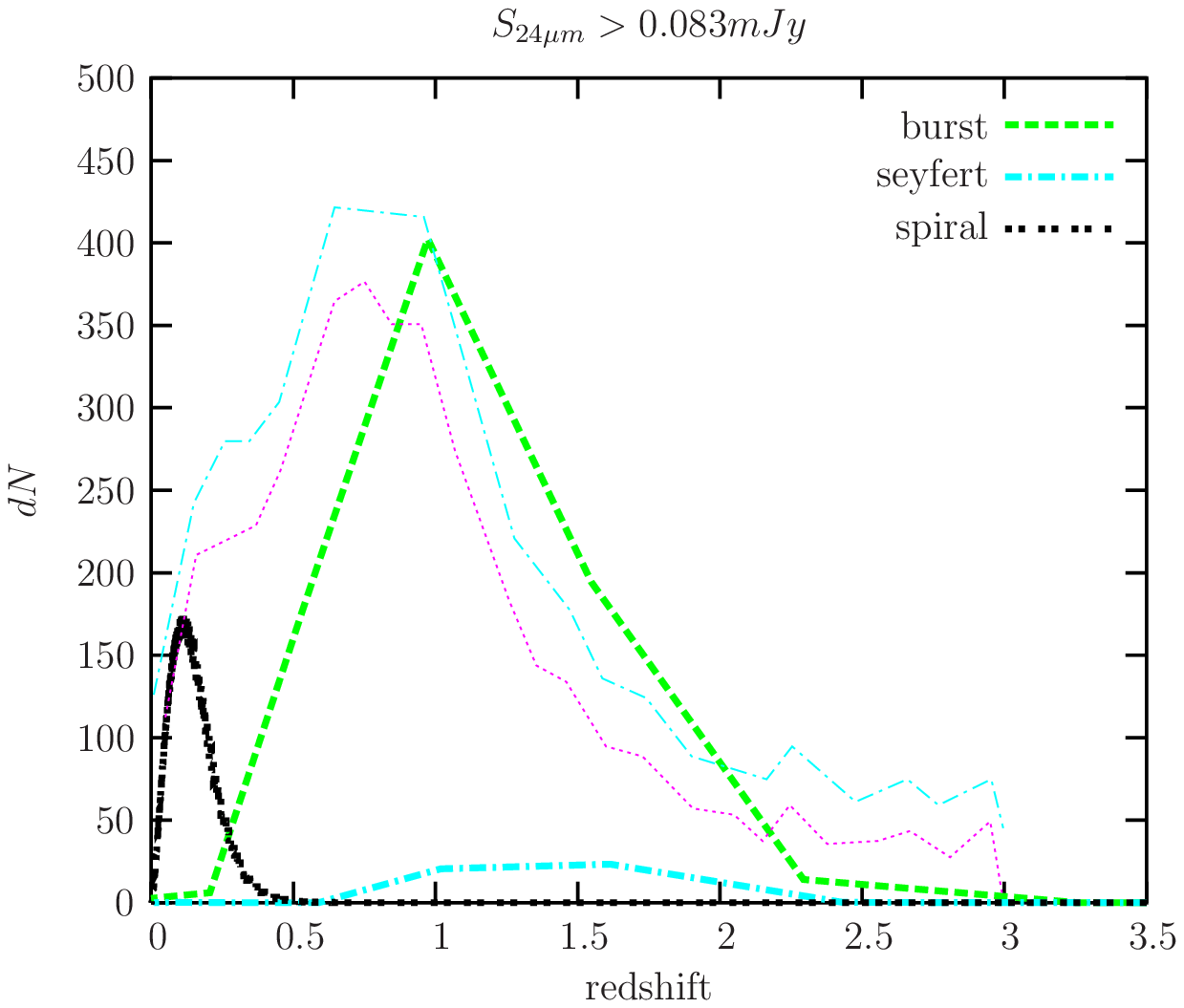}{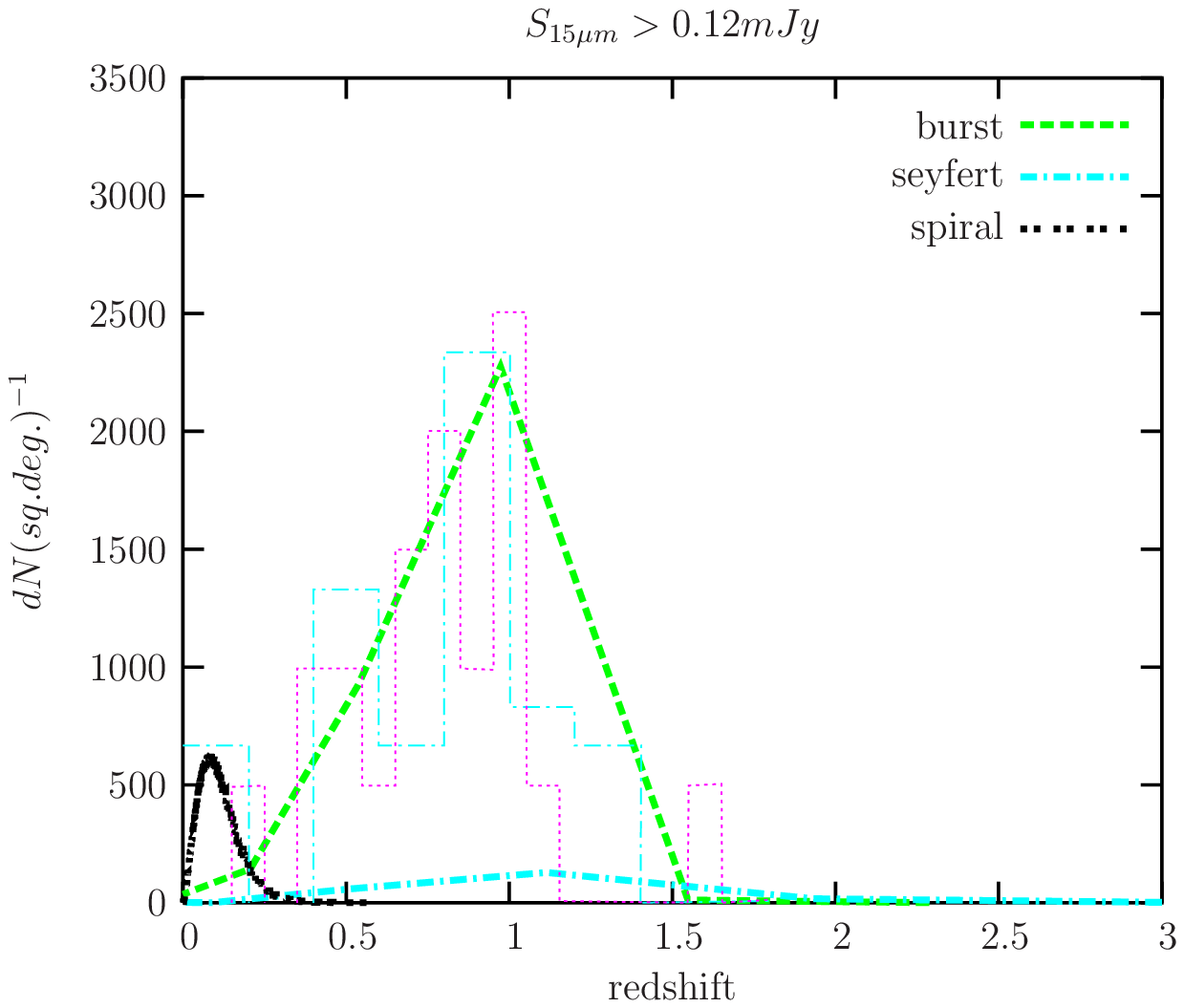}
\caption{Left: The predicted redshift distribution by modelling of three major infrared populations, with a
detection limit $\rm S_{24\mu m}>0.083\,mJy$. A comparison with the recent photometric redshift estimation by
P$\acute{e}$rez-Gonz$\acute{a}$lez et al. (2005) is presented. Blue dot-dashed and red dotted thin lines are
the measured upper and lower limits; Right: The predicted redshift distribution, within a detected limit $>0.12\,mJy$
at $15\,\mu m$. The histogram of the observed redshift distribution of $\rm ISOCAM \, 15\mu m$ sources is taken
from Franceschini et al. (2001). \label{fig9}}
\end{figure}

The redshift distribution of the three model components (starburst galaxies, ``non-evolving'' galaxies
and dusty AGNs) with a detection limit $S_{24\mu m}>0.083\,mJy$ is presented in Fig. \ref{fig9} (Left).
A comparison with the recent photometric redshift estimation by P$\acute{e}$rez-Gonz$\acute{a}$lez et al. (2005)
shows that the starburst population in the modelling can only account for about $30\%$ of the high-z tail ($z>2$).
At these redshifts, the strong PAH peak begins to move out of the MIPS 24$\mu$m band and the strong
mid-infrared continua from AGNs may play an increasing role in the detections. Therefore,
we suspect that dusty AGNs, or the AGN population in a phase with luminosities still
dominated by starbursts, may account for the additional objects in
the high-z tail \citep{Wan03}. We shall in the future properly handle such a population
in expanded modelling.

\begin{figure}
\plottwo{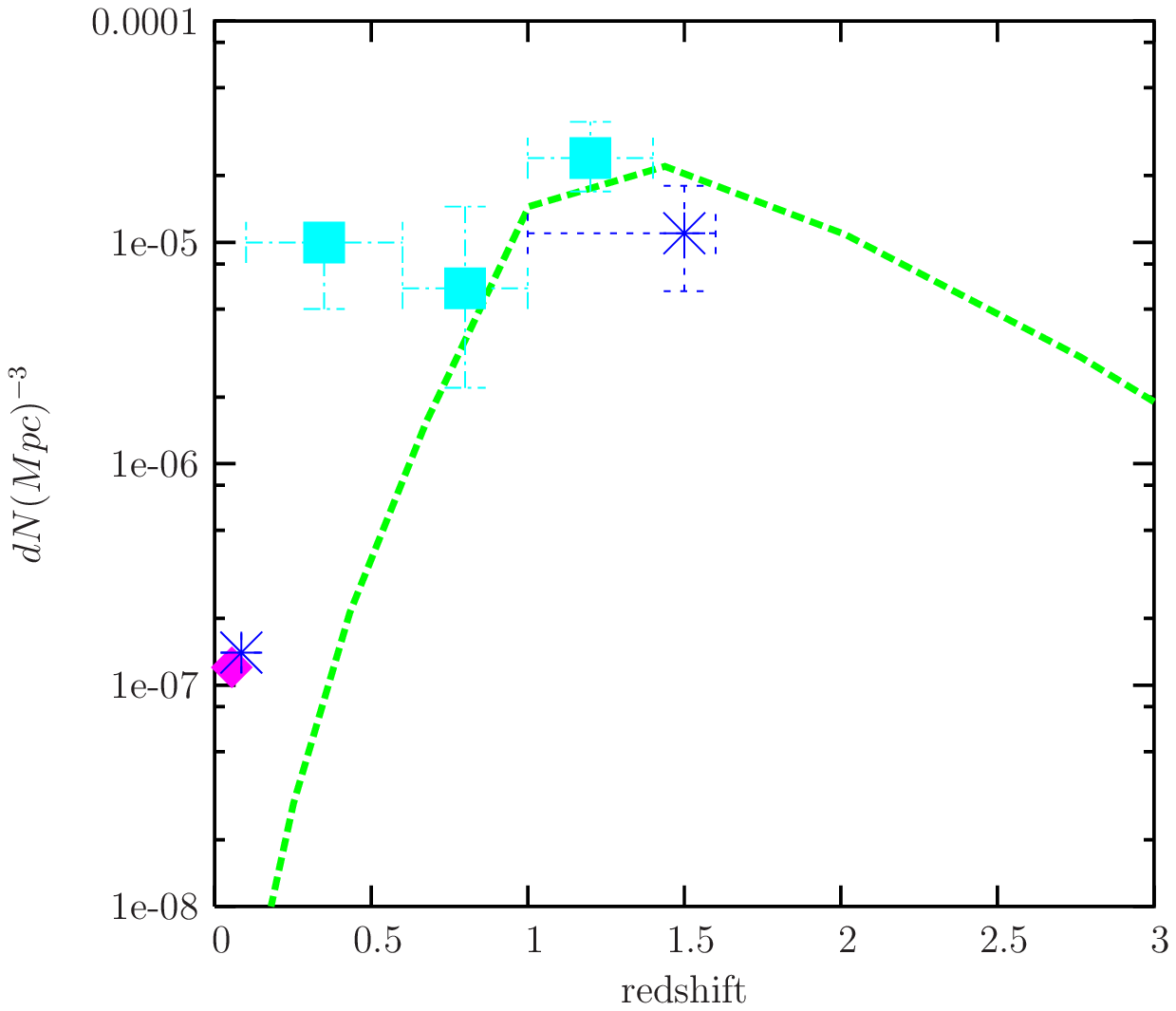}{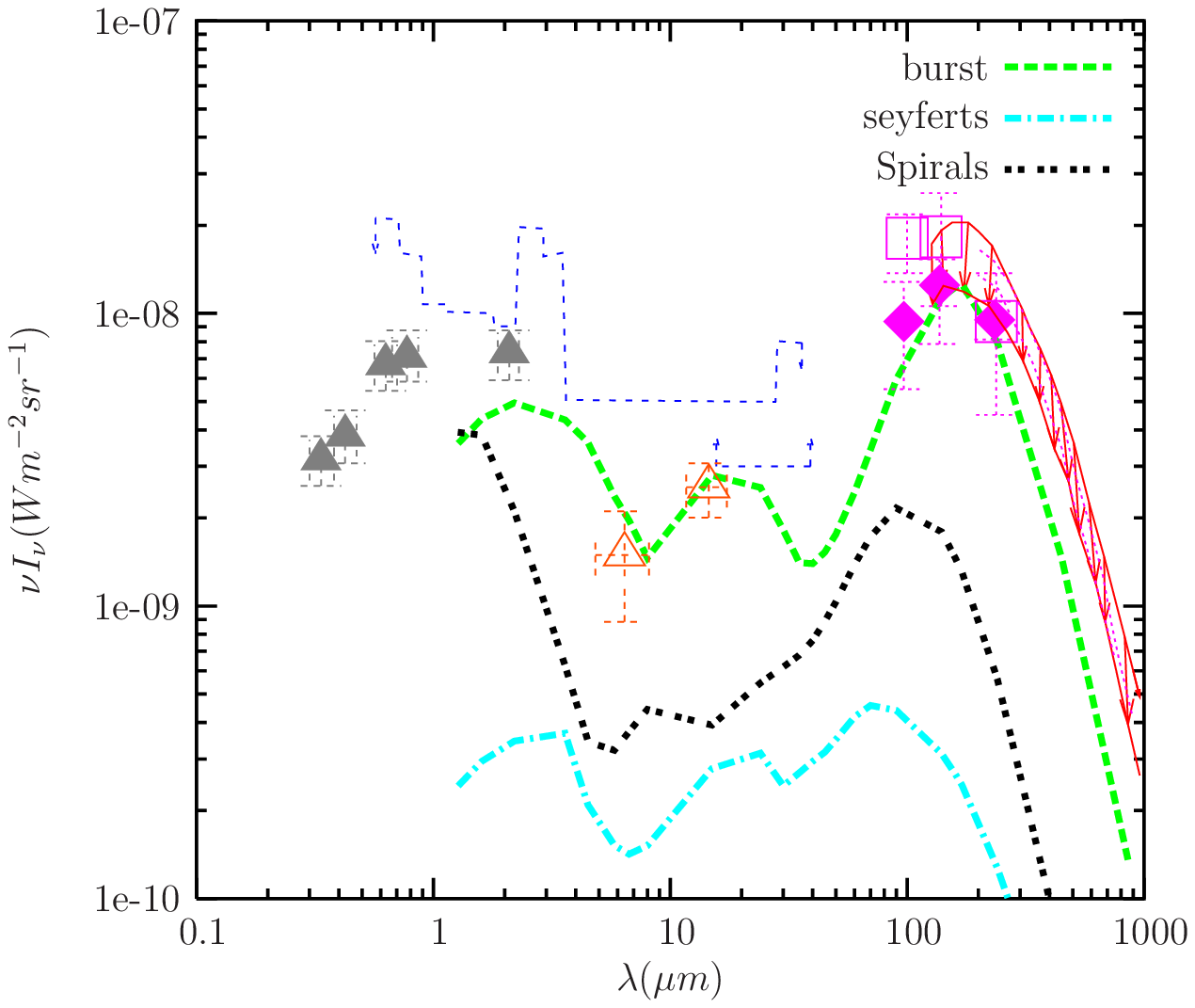}
\caption{Left: The predicted number density distribution of ULIGs (dashed line) is compared with that of the
starburst galaxies with ULIG radio power in the redshift bin $0.1-0.6$, $0.6-1.0$ and $1.0-1.4$ measured by
Cowie et al. 2004 (blue filled squares). The blue asterisks denote the local number density of ULIGs + LIGs
and $\rm z=1-3$ radio/submm sources ($>6\,mJy$) from Barger et al. 2000. The pink diamond shows the estimated
value by Cowie et al. 2004 from the local star forming radio luminosity function; 
Right: Comparison of the calculated background
spectrum with the measurements by independent groups in the all-sky COBE maps (e.g. Hauser et al. 1998), and the
ultradeep optical extragalactic surveys by the HST in the HDF (black filled triangles, Madau \& Pozzetti 2000).
The three lower datapoints (red filled diamonds) in the far-IR are from re-analysis of the DIRBE data by
Lagache et al.(1999), and the red fenced area is upper limits to the far-infrared background level from
Fixsen et al. (1998). The two mid-IR points are the resolved fraction of the CIRB by the deep ISO surveys
IGTES (empty red triangles, Elbaz et al. 2002b), while the dashed
histograms are limits set by TeV cosmic opacity measurements. \label{fig10}}
\end{figure}

The right panel of Fig. \ref{fig9} shows the calculated redshift distribution of the $\rm ISOCAM \,15\mu m$
sources with the detection limit ($>0.12\,mJy$). These luminous infrared sources
cover a wide redshift range of $0.5\sim 2$, peaking at $z\sim
1$. Cowie et al. (2004) recently probed the evolution of the bright tail of the radio luminosity
function up to $z = 1.5$ with redshift measurements of two deep
$\rm 1.4\,GHz$ fields. They found that the number density of the
sources with ULIG radio power and no clear AGN signatures evolves as $(1+z)^7$
up to about redshift $z\sim 1$. A comparison of the $\rm 1.4\,GHz$ radio surveys with our model prediction on
the redshift distribution of ULIGs is shown in the left panel of Fig. \ref{fig10}. We see that
the number density of ULIGs increases dramatically up to redshift $z\sim 1$, but decreases afterwards. We know that
the number density of starburst sources at $z>>1$ is still loosely constrained in the model, and recent detections
of massive galaxies
in the young Universe challenge most of the galaxy evolution theories. However, this part of discussion is out of the
scope
of our current work \citep{Cim04, Gla04}.

Finally, the infrared background from the model calculation is presented in Fig. \ref{fig10} (Right).
Our predictions are consistent
with the direct background measurements from
COBE and other deep infrared surveys, as well as the upper limits from high
energy TeV $\gamma$ ray
detections of nearby Blazars \citep{Fun98, Guy00, Hau01}.

\section{Summary}
We have described a galaxy evolutionary scenario where starburst and
AGN activities are
triggered by galactic mergers/interactions. Assuming an infrared burst phase from gas rich mergers at
high redshift, we have successfully interpreted the available
observations ranging from
near-infrared up to radio deep surveys, leaving the infrared background consistent
with the upper limits from direct infrared background measurements
by COBE and other infrared deep surveys, as well as
recent TeV $\gamma$ ray detections of nearby Blazars.

We further found from the current deep extragalactic surveys performed in the
far-infrared with $\it Spitzer$, and
at radio wavelengths with the VLA, that:

\noindent
1.) LIGs with M82-like enhanced PAH emission features dominate the number counts at sub-mJy flux levels in
$\it Spitzer$ $24\mu m$ and ISOCAM $15\mu m$ deep surveys;

\noindent
2.) the $\rm 1.4\, GHz$ source excess at sub-mJy levels and the strong evolution detected by ISOCAM $15\mu m$ and
$\it Spitzer$ $24\mu m$ extragalactic deep surveys at flux $S_{\nu}< 1\,mJy$ are due to the appearance of the same
population, which is mostly a LIG phase triggered by galaxy interactions at $z\sim 1$; and

\noindent
3.) considering near- and mid-infrared deep surveys and the
background measurements, dusty starburst galaxies triggered by galactic interactions at $z\sim 1$ in the model
account for a moderate fraction of the stellar mass in the Universe ($30\%$ to $50\%$ as an upper limit).

Although our modelling fits the full suite of available data well, it is not likely to be unique.
To break the degeneracies in the model and make further progress requires both a better
understanding of the infrared burst phase, and more accurate merger rate statistics.
With the existing model constraints such as number counts,
redshift distributions and background intensities, it is still very difficult to address such issues in
detail and
give an independent estimation of the model parameters. A real physical model including
both proper chemical evolution and dust emission
properties should be used in future work, for a better understanding of both the emission properties and
dynamical processes of the dusty starburst merging galaxies across cosmic time.

\section*{Acknowledgments}

This work is supported by National Scientific Fundation of China
(NSFC 10173025) and the Chinese 973 project(TG 2000077602). YPW
feels very grateful to Prof. G. Rieke for the helpful discussion and the careful check
of the presentation (his work is partially supported by contract 960785 issued
by JPL/Caltech). YPW would thank the anonymous referee for her/his very kind and
helpful comments, and colleagues at NAOC
for their hospitality.

\clearpage

%%%%%%%%%%%%%%%%%%%%%%% Include Figures, Captions

\end{document}